\documentclass[12pt]{elsarticle}

\usepackage[T1]{fontenc}
\graphicspath{ {./figures/} }
\usepackage{url}
\usepackage{graphicx}
\usepackage{floatrow}
\usepackage{subfig}
\usepackage{xcolor}
\usepackage{tabularx}
\usepackage{booktabs} 
\usepackage{etoolbox}
\usepackage{setspace}
\usepackage{balance}
\newcounter{rowcntr}[table]
\renewcommand{\therowcntr}{\arabic{rowcntr}}
\newcolumntype{N}{>{\refstepcounter{rowcntr}}r}
\AtBeginEnvironment{tabular}{\setcounter{rowcntr}{0}}
\newcommand{\rowref}[2]{{#1-\ref*{#2}}}
\usepackage{dcolumn}
\usepackage{siunitx}
\usepackage{booktabs}
\usepackage[version=4]{mhchem}
\usepackage{collcell}
\newcolumntype{P}{>{\collectcell\num}r<{\endcollectcell}@{${}\pm{}$}>{\collectcell\num}r<{\endcollectcell}}

\usepackage{xparse}   
\newcolumntype{Y}{>{\centering\arraybackslash}X}
\NewDocumentCommand{\rot}{O{45} O{1em} m}{\makebox[#2][l]{\rotatebox{#1}{#3}}}%

\usepackage{amssymb}
\usepackage{pifont}
\usepackage{bbm}
\newcommand{\cmark}{\ding{51}}
\newcommand{\xmark}{\ding{53}}

\usepackage{lineno,hyperref}
\modulolinenumbers[5]

\usepackage[utf8]{inputenc}
\usepackage{amssymb}
\usepackage{amsmath}
\usepackage{enumitem}
\usepackage[ruled,vlined,linesnumbered]{algorithm2e}
\usepackage{booktabs}

\newcommand{\abs}[1]{\lvert#1\rvert}
\newcommand{\sat}[1]{\text{sat}#1}
\newcommand{\pp}[3]{#1 \left(#2 \middle| #3\right)} 

\usepackage[capitalise, noabbrev]{cleveref}

\journal{arXiv}

\begin{document}

\begin{frontmatter}


\title{Deep Reinforcement Learning with Shallow Controllers: An Experimental Application to PID Tuning}


 \author[math]{Nathan P. Lawrence\corref{nathan}}
 \cortext[nathan]{input@nplawrence.com}
 \author[honeywell]{Michael G. Forbes}
 \author[math]{Philip D. Loewen}
 \author[chbe]{Daniel G. McClement}
 \author[backstrom]{Johan U. Backstr{\"o}m}
 \author[chbe]{R. Bhushan Gopaluni}

\address[math]{Department of Mathematics, University of British Columbia, Vancouver BC, Canada}
\address[chbe]{Department of Chemical and Biological Engineering, University of British Columbia, Vancouver, BC Canada}
\address[honeywell]{Honeywell Process Solutions, North Vancouver, BC Canada}
\address[backstrom]{Backstrom Systems Engineering Ltd.}
\begin{abstract}
Deep reinforcement learning (RL) is an optimization-driven framework for producing control strategies for general dynamical systems without explicit reliance on process models. 
Good results have been reported in simulation. %
Here we demonstrate the challenges in implementing a state of the art deep RL algorithm on a real physical system.
Aspects include the interplay between software and existing hardware; experiment design and sample efficiency; training subject to input constraints; and interpretability of the algorithm and control law.
At the core of our approach is the use of a PID controller as the trainable RL policy.
In addition to its simplicity, this approach has several appealing features: No additional hardware needs to be added to the control system, since a PID controller can easily be implemented through a standard programmable logic controller; the control law can easily be initialized in a ``safe'' region of the parameter space; and the final product---a well-tuned PID controller---has a form that practitioners can reason about and deploy with confidence.
\end{abstract}


\end{frontmatter}


\floatsetup[figure]{style=plain,subcapbesideposition=top}

\section{Introduction}
\label{sec:intro}

Reinforcement learning (RL) is a branch of machine learning that adaptively formulates a ``policy'' through interactions with an environment \citep{suttonReinforcementLearningIntroduction2018}. Such a general framework has led to the burgeoning interest in RL for process control applications \citep{shin2019ReinforcementLearning, nian2020ReviewReinforcement}. However, real-world applications in the field remain sparse. Systems integration and software development, alongside more fundamental algorithmic issues concerning closed-loop stability and sample efficiency are all formidable challenges one faces.\par

We take a pragmatic approach to implementing RL. Proportional-integral-derivative (PID) controllers are well-understood and ubiquitous in practice. This makes the problem of tuning a PID a reasonable starting point for real-world applications of RL. Moreover, their simple structure and stabilizing properties are neatly compatible with policy design in RL. Further, the prevalence of PID controllers in current industrial plants makes it natural to seek RL applications that operate effectively with them and thus take full advantage of existing hardware and expertise. Therefore, this framework provides a fruitful testbed for evaluating RL algorithms through the lens of industrially-accepted auto-tuning methods.\par

This paper extends the previous work of \citet{lawrence2020OptimalPID} to a physical system; the key idea is to interpret a PID controller as the RL policy.
We embrace the fact that performance is not the only metric of interest when evaluating a new approach to controller design \citep{astrom1984AutomaticTuning, forbes2015ModelPredictive}.
Other important considerations include \emph{ease of use}, \emph{maintainability}, and \emph{robustness}.
We convey the technical challenges involved in implementing a RL algorithm in our lab: this includes special consideration of the interplay between software development and existing hardware.
Moreover, we propose various metrics for evaluating RL algorithms with existing auto-tuners serving as a baseline.
We aim to show that RL can be deployed on a physical system in an interpretable and modular fashion, and to weigh the merits of RL against standard tuning techniques.
Ultimately, the purpose of this work is to develop a guide for real-world applications of RL in process control.

\subsection{Related work}
\label{subsec:related}

We review some related work at the intersection of RL and process control.
For a more thorough overview the reader is referred to the survey papers by \citet{shin2019ReinforcementLearning, lee2018MachineLearning}, or the tutorial-style papers by \citet{nian2020ReviewReinforcement, spielberg2019SelfDriving}.
Moreover, since PID control is such a large field, we limit our survey of PID control to RL-related methods; some data-driven and optimization-based approaches can be found in the works of \citet{berner2018ExperimentalComparison, wakitani2019DesignApplication}, and others they cite.

\par

One of the first studies of reinforcement learning for process control applications is due to \citet{hoskins1992ProcessControl}. Later, in the early 2000s, several successful implementations of RL methods in process control were developed. For example, \citet{lee2008ValueFunctionbased, kaisare2003SimulationBased} utilize approximate dynamic programming methods for optimal control of discrete-time nonlinear systems.
These early works demonstrate the applicability of RL in process control through applications such as scheduling problems or control of a microbial cell reactor when a simulation model is available.\par
More recently, there has been significant interest in deep RL methods for process control \citep{noel2014ControlNonlinear, syafiie2011ModelfreeControl, ma2019ContinuousControl, cui2018FactorialKernel, ge2018ApproximateDynamic, pandian2018ControlBioreactor, dogru2021OnlineReinforcement}. \citet{spielberg2019SelfDriving} adapted the popular model-free deep deterministic policy gradient (DDPG) algorithm for setpoint tracking problems.
Meanwhile, \citet{wang2018NovelApproach} developed a deep RL algorithm based on proximal policy optimization algorithm \citep{schulman2017ProximalPolicy}. \citet{petsagkourakis2020ReinforcementLearning} use transfer learning to adapt a policy developed in simulation to novel systems. Variations of DDPG, such as twin-delayed DDPG (TD3) \citep{fujimoto2018AddressingFunction} or a Monte-Carlo based strategy have also shown promising results in complex control tasks \cite{joshi2021ApplicationTwin, yoo2021ReinforcementLearninga}. Other approaches utilize meta-learning and apprenticeship learning to quickly adapt trained RL models to new processes \citep{mcclement2021MetaReinforcementLearning, mowbray2021UsingProcess}.
In the model-based setting, \citet{kim2020ModelbasedDeep} incorporate deep neural networks (DNNs) as value function approximators into the globalized dual heuristic programming algorithm. Predictive models have also been augmented with popular DRL algorithms, such as DDPG or TD3, to improve the policy gradient estimation \citep{bao2021DeepReinforcement}
\par
Other approaches to RL-based control postulate a fixed control structure such as PID \citep{sedighizadeh2008AdaptivePID, shipman2019ReinforcementLearning, carlucho2017IncrementalLearning}.
\citet{brujeni2010DynamicTuning} develop a model-free algorithm to dynamically select the PID gains from a pre-defined collection derived from Internal Model Control (IMC). Their approach is adapted to a physical continuously stirred tank heater after pre-training in simulation.
\citet{berger2013NeurodynamicProgramming} dynamically tune a PID controller in continuous parameter space using the actor-critic method; their approach is based on dual heuristic dynamic programming, where an identified model is assumed to be available. The actor-critic method is also employed by \citet{sedighizadeh2008AdaptivePID}, where the action at each time-step is to revise the PID gains.\par 
Only a handful of these studies on RL for process control contain real-world validation \citep{dogru2021OnlineReinforcement, brujeni2010DynamicTuning, pandian2018ControlBioreactor, syafiie2011ModelfreeControl, nian2020ReviewReinforcement}. This work is focused on evaluating the merits of deep RL as a model-free auto-tuning strategy. This entails a couple significant differences from previous work: We do not use a simulation model or offline datasets for pre-training the RL agent; We outline criteria for any auto-tuning strategy and evaluate our deep RL algorithm through this lens. Crucially, we provide a detailed account of the hardware, software, and algorithmic details involved in implementing the RL agent.

\section{Methodology}
\label{sec:methodology}

\begin{figure}[tbh]
\begin{center}
\includegraphics[width=\textwidth]{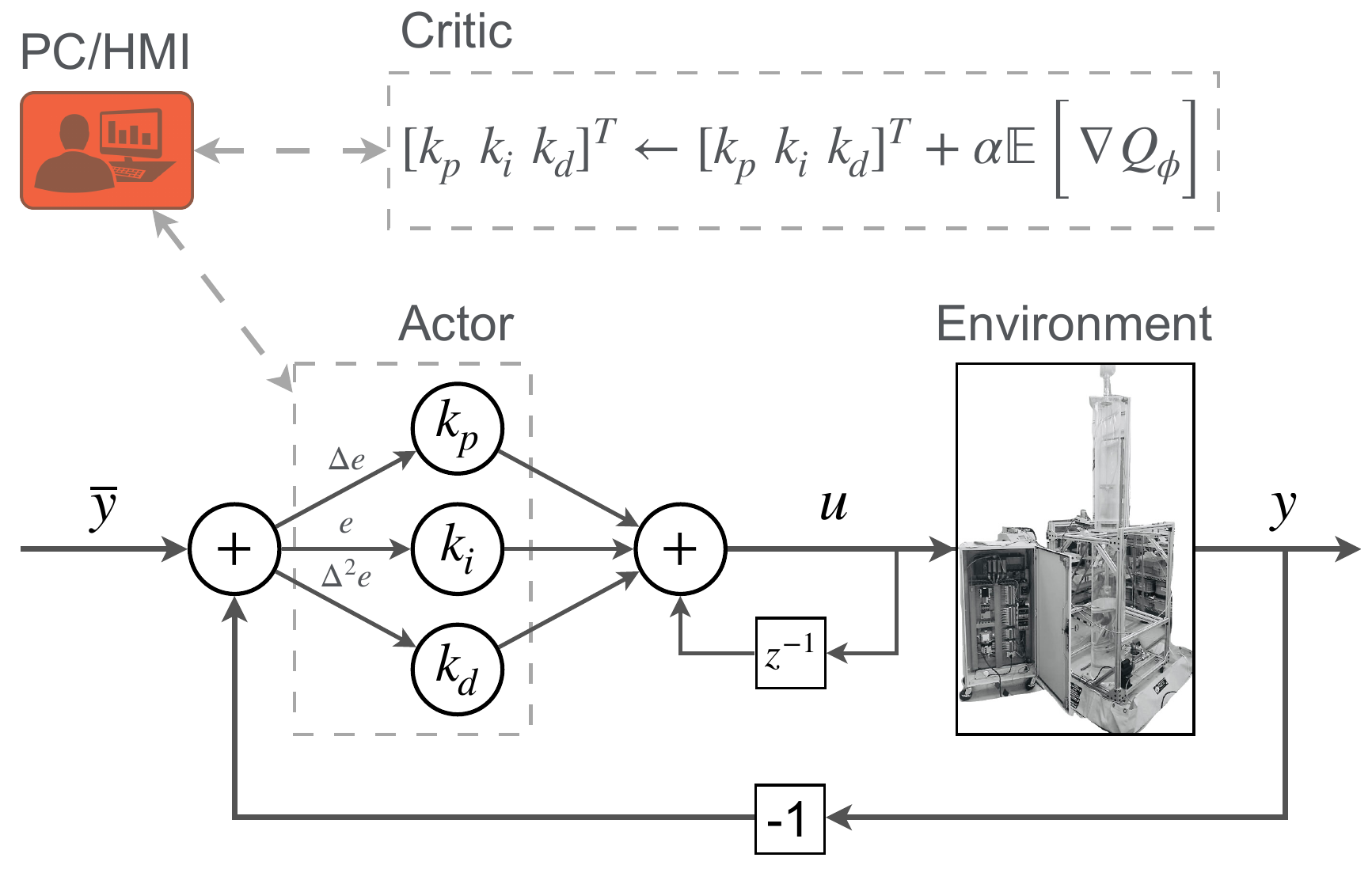}
\caption{A conceptual diagram of the proposed method. The actor network is a PID controller, which interacts with a physical two-tank system. The PC/HMI collect process data, which is used by the RL algorithm to train a critic network. The critic network then updates the PID parameters, which are read by the HMI and sent back to the system. Dashed lines indicate a separation of time scales between the RL algorithm and the physical system.}
\label{fig:conceptual}
\end{center}
\end{figure}

\subsection{Reinforcement learning}
\label{subsec:RLbackground}

A brief overview of deep RL will serve to fix our notation, which is largely standard.
For more background, see \citet{suttonReinforcementLearningIntroduction2018}; tutorial-style treatments are given by \citet{nian2020ReviewReinforcement, spielberg2019SelfDriving}.

\par

The system of interest has states $s$ that evolve in some set $\mathcal{S}$.
At each instant $t$ (an integer), the agent observes the state $s_t$ and responds by applying some action $a_t$, chosen from a given set $\mathcal{A}$.
The system dynamics then produce a new state $s_{t+1}$,and the agent receives a scalar reward, $r_t$.
As time marches on, a history $h = (s_0, a_0, s_1, a_1, \ldots)$ is produced.
The present value of the agent's total accumulated rewards is
\[
\sum_{t=1}^\infty \gamma^{t-1} r_t,
\]
where $\gamma\in(0,1]$ is a fixed discount factor.

\par

Randomness complicates the situation.
Given the current state $s_t$ and action $a_t$, the new state $s_{t+1}$ is a random variable distributed according to some density $\pp{p}{\cdot}{s_t,a_t}$.
(Typically $p$, which encodes the system's nonlinear stochastic dynamics, is taken as completely unknown.)
More formally, the system is assumed to be a Markov Decision Process (MDP).
The agent's reward emerges from a fixed function $r$, according to $r_t=r(s_t,a_t)$.
The agent's actions are determined by selecting a ``policy'' $\pi$, which is a state-dependent conditional probability density on $\mathcal{A}$, so that the agent's actions obey $a_t \sim \pp{\pi}{\cdot}{s_t}$.

\par

By fixing a policy $\pi$, the agent completes the specification of a Markov process and establishes a probability density $p^\pi(\cdot)$ on the set of trajectories $h=(s_0,a_0,s_1,a_1,\ldots)$.
The agent's goal is to maximize expected long-term reward\footnote{Standard RL terminology calls for \emph{maximizing a reward}. The problem can be recast as \emph{minimizing a loss} by changing the sign of the objective. Thus $-r_t$ has a role analogous to the stage cost in MPC; when we discuss a reward function with reference to a cost function, this change of sign is understood implicitly.}
by choosing the best possible policy $\pi$, i.e.,
\begin{equation}
\begin{aligned}
    &\text{maximize} && J(\pi) = \mathbb{E}_{h \sim p^{\pi}(\cdot)}\left[ \sum_{t=1}^{\infty} \gamma^{t-1}r(s_t,\pi(s_t)) \middle| s_0 \right]\\
    &\text{over all} && \text{policies } \pi \colon \mathcal{S} \to \mathcal{P}(\mathcal{A}),
\end{aligned}
\label{eq:RLobjective}
\end{equation}
where $\mathcal{P}(\mathcal{A})$ denotes the set of probability measures on $\mathcal{A}$.\par

The broad subject of reinforcement learning concerns iterative methods for choosing a desirable policy $\pi$ (this is the ``learning''), guided in some fundamental way by the agent's observations of the rewards from past state-action pairs (this provides the ``reinforcement'').

We next outline the class of algorithms aimed at solving Problem~\eqref{eq:RLobjective}.
Not knowing $p$ (or, consequently, $p^\pi$) is a key limitation.
We focus exclusively on methods that produce \emph{deterministic} policies $\mu$, which can be considered as simple $\mathcal{A}$-valued mappings defined on the state space. Henceforth we use the notation $\mu$ for a policy; in the context of the above formulation one may set $\pi=\mu$.


Common approaches to solving Problem~\eqref{eq:RLobjective} involve $Q$-learning (value-based methods) and the policy gradient theorem (policy-based methods) \citep{suttonReinforcementLearningIntroduction2018}.
These methods form the basis for deep RL algorithms, that is, algorithms for solving RL tasks with the aid of deep neural networks.
Deep neural networks act as flexible function approximators, well-suited for expressing complex control laws.
Moreover, function approximation methods make RL problems tractable in continuous state and action spaces \citep{lillicrap2015ContinuousControl, silver2014DeterministicPolicy, sutton1999PolicyGradient}. Without them, discretization of the state and action spaces is necessary, accentuating the ``curse of dimensionality''.\par

In the space of all possible policies, the optimization is performed over a subset parametrized by some vector $\theta$. For example, in some applications, $\theta$ denotes the set of all weights in a deep neural network. In this work, we take $\theta$ to be the gains in a PID controller.
The policies considered are denoted $\mu_{\theta}$, and we simplify Problem~\eqref{eq:RLobjective} by writing $J(\theta)$ instead of $J(\mu_\theta)$ and $p^\theta$ instead of $p^{\mu_\theta}$.

A standard approach to solving Problem~\eqref{eq:RLobjective} uses gradient ascent:
\begin{equation}
\theta
\leftarrow \theta + \alpha\nabla J(\theta),
\label{eq:PolicyGradient_Iteration}
\end{equation}
where $\alpha > 0$ is a step-size parameter.
Analytic expressions for $\nabla J(\theta)$ exist for both stochastic and deterministic policies \citep{suttonReinforcementLearningIntroduction2018, silver2014DeterministicPolicy}.
Crucially, these formulas rely on the state-action value function, 
\begin{equation}
    Q(s_t, a_t) = \mathbb{E}_{h \sim p^\theta(\cdot)}\left[ \sum_{k = t}^{\infty} \gamma^{k-1}r(s_k,\mu_{\theta}(s_k)) \middle| s_t, a_t \right].
\label{eq:Qfunc}
\end{equation}
Although $Q$ is not known precisely, as it depends on both the dynamics and the policy, it can be estimated with a deep neural network \citep{mnih2015HumanlevelControl}.
Writing $\phi$ for the vector of parameters in this network, and $Q_\phi$ for the approximation to $Q$ that it defines, $\phi$ is chosen to minimize the temporal difference error across $N$  observations indexed by $i$ (or variations of this, as given in the references cited below):
\begin{equation}
    \mathcal{L}_{\text{critic}}(\phi) = \frac{1}{N} \sum_{i = 1}^{N} \left(Q_{\text{target}}^{(i)} - Q_{\phi} (s^{(i)}, a^{(i)}) \right)^2.
\end{equation}
For example, $Q_{\text{target}}^{(i)} = r(s^{(i)}, a^{(i)}) + \gamma Q_{\phi} (s'^{(i)}, \mu_{\theta} (s'^{(i)}))$, and $s'$ represents the next state in the trajectory following policy $\mu_{\theta}$.
With an up-to-date critic network, we then define the loss for the actor network as follows: 
\begin{equation}
    \mathcal{L}_{\text{actor}}(\theta)
    = \frac{1}{N} \sum_{i = 1}^{N} Q_{\phi}(s^{(i)}, \mu_{\theta}(s^{(i)})).
    \label{eq:actorloss}
\end{equation}
$\nabla \mathcal{L}_{\text{actor}}$ serves as a tractable approximation of $\nabla J$, and is therefore used in the nominal policy update shown in \cref{eq:PolicyGradient_Iteration} \citep{silver2014DeterministicPolicy}.
These ideas are the basis of popular deep RL algorithms such as DDPG, TD3, SAC \citep{lillicrap2015ContinuousControl, fujimoto2018AddressingFunction, haarnoja2018SoftActorCritic}. More generally, they fall into the class of \emph{actor-critic} methods \citep{konda2000ActorcriticAlgorithms}, as they learn both a policy $\mu$ ($\approx\mu_{\theta}$) and a state-action value function $Q$ ($\approx Q_{\phi}$).
We use a modified version of TD3, the twin-delayed DDPG algorithm \citep{fujimoto2018AddressingFunction}, a refinement of DDPG \citep{lillicrap2015ContinuousControl}.
For completeness, we present the basic algorithm in \cref{alg:ActorCriticProposed} and give an overview of its main differences from DDPG in \ref{app:details}.

\subsection{PID in the RL framework}
\label{subsec:PIDRL}

We now apply the general formulation given in the previous section to the problem of PID tuning. The proposed framework is illustrated in \cref{fig:conceptual}. We use the variable $\bar{y}$ to denote a reference signal; we generally omit the time index with the understanding that the initial time $t=0$ indicates a step change in the reference signal, which is then constant. We then write the output error signal as $e^{(y)}_t = \bar{y} - y_t$. We use the operator $\Delta$ to denote a first-order difference between time steps of a signal; for example, $\Delta e^{(y)}_t = e^{(y)}_t - e^{(y)}_{t-1}$. We use $\Delta t$ to denote the sampling time. 
When implementing a PID controller in the incremental form, the derivative term requires second-order output information. We use a superscript $f$ to denote a signal resulting from a low-pass filter; since the only instances in which we use this convention are to approximate derivatives, we include the sampling time in the definition. The filtered first-order difference of the output $\Delta y^{(f)}_{t}$ is used to compute a second-order difference, based on the following recursion:
\begin{align}
\Delta y^{(f)}_{t} &= T_f \Delta y^{(f)}_{t-1} + (1-T_f)\frac{y_t - y_{t-1}}{\Delta t}\\
\Delta^2 y_{t}^{(f)} &= \frac{\Delta y^{(f)}_{t} - \Delta y^{(f)}_{t-1}}{\Delta t},
\label{eq:derivTerm}
\end{align}
where $T_f \in [0,1]$ is a fixed parameter.
An anti-windup component\footnote{Although we are working with the incremental form of PID controller, which automatically resets the controller at its saturation limits, we still use the terminology ``anti-windup'' to refer to this component of the controller.} is incorporated through the variable $e^{(u)}_{t-1} = \hat{u}_{t-1} - u_{t-1}$, where $\hat{u}$ is the signal proposed by a PID controller before being saturated and $u = \text{sat}(\hat{u})$ denotes the actual input signal after saturation.\par
With the variables $\Delta e^{(y)}_{t}$, $e^{(y)}_{t}$, $\Delta^2 y_{t}^{(f)}$, $e^{(u)}_{t-1}$, $u_{t-1}$, 

a PID controller can then be written as follows:
\begin{align}
\begin{split}
    \hat{u}_{t} =&
    \begin{bmatrix}
    k_p & k_i & k_d & k_{\tau}
    \end{bmatrix}
    \begin{bmatrix}
    \Delta e^{(y)}_{t}\\
    \Delta t e^{(y)}_{t}\\
    -\Delta^2 y^{(f)}_{t}\\
    \Delta t e^{(u)}_{t-1}
    \end{bmatrix} + u_{t-1}.
    \label{eq:SS2}
\end{split}
\end{align}
The scalar $\hat u_t$ given by \cref{eq:SS2} is clipped to produce the physically admissible input signal $u_{t} = \text{sat}\big(\hat{u}_{t}\big)$, which is sent to the plant.
Since \cref{eq:SS2} contains first and second-order output information, and the system usually contains time delay, the RL state is taken to be a history of the observations $o_t = \left[\Delta e^{(y)}_{t}, \Delta t e^{(y)}_{t}, -\Delta^2 y^{(f)}_{t}, \Delta t e^{(u)}_{t-1}, u_{t-1}\right]$:
\begin{equation}
s_{t} = \left[o_{t-d}, \ldots, o_{t}\right]
\label{eq:RLstate}
\end{equation}

where $d$ is a non-negative integer. In the context of the RL formulation in \cref{subsec:RLbackground}, the ``actions'' $a_t$ are interchangeable with the control inputs $u_t$. \par
\cref{eq:RLstate} contains the necessary information for implementing a PID controller in discrete time steps. \cref{eq:SS2} parameterizes the PID controller. We therefore take \cref{eq:SS2} to be a shallow neural network, where $[k_p\ k_i\ k_d\ k_{\tau}]$ is a vector of trainable weights. In light of \cref{eq:PolicyGradient_Iteration,eq:actorloss,eq:SS2}, 
the PID update equation takes the following explicit form:
\begin{align}
	\begin{split}
	\theta &\leftarrow \theta + \alpha\nabla J(\theta)\\
	\end{split}\\
	\begin{split}
	&\approx \theta + \alpha \nabla \left( \frac{1}{N} \sum_{i = 1}^{N} Q_{\phi}(s^{(i)}, \mu_{\theta}(s^{(i)}))\right)\label{eq:actorgradient}\\
	\end{split}\\
	\begin{split} &= [k_p\ k_i\ k_d\ k_{\tau}]^T +\ldots \\ 
	&\ \alpha \frac{1}{N} \sum_{i=1}^{N} \left. \left(\left[\Delta e^{(y)}, \Delta t e^{(y)}, -\Delta^2 y^{(f)}, \Delta t e^{(u)}\right]^T\right)^{(i)} \nabla_{a} Q_{\phi}(s^{(i)}, a)\right|_{a = \mu_{\theta}(s^{(i)})}.\label{eq:explicitactorgradient}	
	\end{split}
\end{align}
\cref{eq:explicitactorgradient} follows by the chain rule and can be computed automatically in deep learning frameworks, such as PyTorch, based on the computation of $\mathcal{L}_{\text{actor}}$ and $\mathcal{L}_{\text{critic}}$.
Later in \cref{subsec:inputConstraint} we explain strategies for training the weights subject to the saturation nonlinearity.

\section{Scorecard for RL algorithms in process control}
\label{sec:scorecard}

We propose the following  criteria for evaluating RL based tuning methods. 

\begin{enumerate}[leftmargin=*]
    \item \emph{Nominal performance}: How much does the RL agent improve the performance of the closed-loop system compared to its initialization?
    \item \emph{Stability}: Is the trained closed-loop system stable? Did it become unstable during training?
    \item \emph{Perturbation to the system}: How much perturbation to the process is required to achieve satisfactory performance?
    \item \emph{Initialization}: To what extent does the performance of the RL agent depend on the initial PID parameters?
    \item \emph{Hyperparameters}: Do the algorithm hyperparameters need to be adjusted between experiments? Which hyperparameters influence the learning process most strongly?
    \item \emph{Training duration}: How many episodes are required to achieve satisfactory performance?
    \item \emph{Practicality and specialization}: What hardware is required in order to deploy the RL algorithm on the physical process? What level of user expertise is required in order to implement the algorithm?
\end{enumerate}

Many of these criteria have natural counterparts in current tuning methods, which provide useful points of reference in our evaluation. These are discussed further in \cref{subsec:evaluation}. 
To quantify our assessment, we consider the integral absolute error (IAE), integral squared error (ISE), total variation (TV), overshoot (OS), and settling time (ST):
\begin{equation}
\begin{aligned}
    &\text{IAE:}\; \int_{0}^{\infty} \abs{e(t)} dt &&\text{ISE:}\; \int_{0}^{\infty} e(t)^2 dt\\
    &\text{TV:}\; \sum_{t=0}^{\infty} \abs{y(t+1) - y(t)} &&\text{OS:}\; \max_{0\leq t < \infty}\left\{ \abs{e(t)} \mathbbm{1}_{\{t \colon e(0) \cdot e(t) < 0 \}}(t) \right\}\\
    &\text{ST:}\; \min_{0\leq t < \infty} t \mathbbm{1}_{\{t \colon \abs{e(\tau)} \leq \epsilon\; \forall \tau \geq t \}}(t) &&\text{where}\;
    \mathbbm{1}_{\Omega}(t) =
    \begin{cases}
    1\quad \text{if}\; t \in \Omega\\
    0\quad \text{otherwise}.
    \end{cases}
\end{aligned}
\label{eq:metrics}
\end{equation}
Of course, these are approximated as summations using discrete-time samples from the system. To compute percent overshoot (\% OS), the OS is multiplied by $100/\abs{\Delta y_{\text{sp}}(0)}$. We will also be interested in the total variation of the input signal, which is denoted by $\text{TV}_u$.\par

These criteria address much more than simply the performance of the trained RL agent. We are interested in the ``path'' the RL agent takes to reach its final form, the perturbations it makes to the system, its usability, and the overall software/hardware requirements.

\section{Lab setup and software structure}
\label{sec:experimentation}

\begin{figure}[tbh]
\begin{center}
\includegraphics[width=0.5\linewidth]{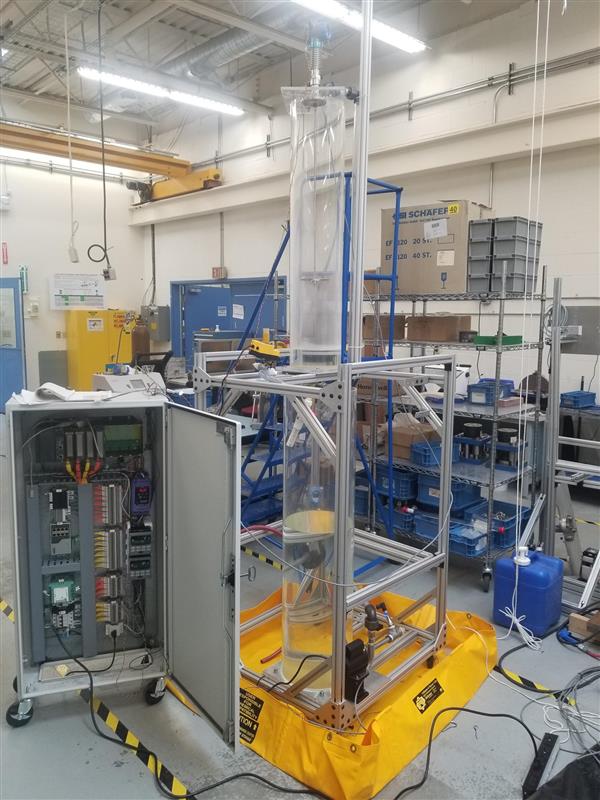}
\caption{Our two-tank system in a lab at Honeywell for testing PID tuning algorithms.}
\label{fig:TwoTank}
\end{center}
\end{figure}

In this section, we describe the dynamics and instrumentation of our two-tank system. We also give an overview of the software used to implement a RL agent on the physical system with standard hardware.

\subsection{Description of the two-tank system}
\label{subsec:dynamics}

We consider the problem of controlling the liquid level in a tank, using the physical apparatus shown in \cref{fig:TwoTank}. The tank of interest is positioned vertically above a second tank that serves as a reservoir. Water drains from the tank into the reservoir through an outflow pipe, and is replenished by water from the reservoir delivered by a pump whose flow rate is our manipulated variable. More precisely, two PID controllers are in action: For a desired level, one PID controller outputs the desired flow rate based on level tracking error. This flow rate is then used as a reference signal for the second PID controller, whose output is the pump speed. The first is referred to as the ``level controller'' and the second as the ``flow controller''.\par
Significant physical dimensions (with units of length) include
$r_{\text{tank}}$, the radius of the top tank;
$r_{\text{pipe}}$, the radius of the pipe for the outflow of the top tank; and
$\ell$, the distance from the base of the top tank up to the water surface.
(We later refer to $m$, a filtered counterpart of $\ell$.)

Key flow parameters, with units of volume/time,
are the outflow $f_{\text{out}}$;
the inflow $f_{\text{in}}$;
an empirical coefficient $f_c$;
and $f_{\text{max}}$, the maximum flow the pump can deliver.\par
The pump delivers water to the top tank at the rate
$(p/100)f_{\rm max}$,
where $p$ is a dimensionless percentage in $[0,100]$.

We write $\bar{p}$ for the commanded pump speed,
typically a piecewise-constant setpoint function.

The system dynamics are based on Bernoulli's equation, $f_{\text{out}}\approx f_c\sqrt{2 g\ell}$, and the conservation of fluid volume in the upper tank:

\begin{equation}
\frac{d\hfil}{dt}\left(\pi r_{\text{tank}}^2\ell\right)
= \pi r_{\text{tank}}^{2} \dot{\ell}
= f_{\text{in}} - f_{\text{out}}.
\end{equation}
(We use a dot for the time derivative; $g$ is the gravitational constant.)

Our system model combines the principles above with some simple filters to make our mathematical description physically realizable. A first-order filter with time constant $\tau\ge0$
transforms an input signal $\hat{y}$
into the output signal $y$ defined by
\begin{equation}
\tau \dot{y} + y = \hat{y},
\qquad y(0)=0.
\end{equation}
Note that setting $\tau=0$ gives $y(t)=\hat y(t)$, while if $\hat y$ is constant,
one has $y(t) = (1 - e^{-t/\tau})\hat y$.
Our application involves four filtered signals, with time constants
$\tau_p$ for the pump,
$\tau_{\text{in}}$ for changes in the inflow,
$\tau_{\text{out}}$ for the outflow,  and
$\tau_m$ for the measured level dynamics.

We therefore have the following system of differential equations describing the pump, flows, level, and measured level:
\begin{align}
\tau_p \dot{p} + p &= \bar{p}\label{eq:pump}\\
\tau_{\text{in}} \dot{f}_{\text{in}} + f_{\text{in}} &= f_{\text{max}} \left(\frac{p}{100}\right)\label{eq:fin}\\
\tau_{\text{out}} \dot{f}_{\text{out}}+ f_{\text{out}} &= \pi r_{\text{pipe}}^{2} f_{\text{c}} \sqrt{2 g \ell}\label{eq:fout}\\
\pi r_{\text{tank}}^{2} \dot{\ell} &= f_{\text{in}} -  f_{\text{out}}\label{eq:level}\\
\tau_m \dot{m} + m &= \ell. \label{eq:measure}
\end{align}
To track a desired level $\bar{\ell}$, we can employ level and flow controllers by including the following equations:
\begin{align}
\bar{p} &= \text{PID}_{\text{flow}} (\bar{f}_{\text{in}} - f_{\text{in}})\label{eq:PIDflow}\\
\bar{f}_{\text{in}} &= \text{PID}_{\text{level}}(\bar{\ell} - m)\label{eq:PIDlevel}.
\end{align}
\cref{eq:PIDflow}--\eqref{eq:PIDlevel} use shorthand for PID controllers taking the error signals $\bar{f}_{\text{in}} - f_{\text{in}}$ and $\bar{\ell} - m$, respectively. For our purposes, $\text{PID}_{\text{flow}}$ is fixed and a part of the environment, while $\text{PID}_{\text{level}}$ is the tunable controller.

This mathematical description is given to provide intuition for our control system. Moreover, it was used for the offline development of our interface for interacting with the physical system.

\subsection{Human-machine interface}
\label{subsec:HMI}
\begin{figure}[!phtb]
\begin{minipage}[]{\linewidth}
\begin{center}
    \sidesubfloat[]{%
        \includegraphics[width=\linewidth, height=0.45\textheight, keepaspectratio]{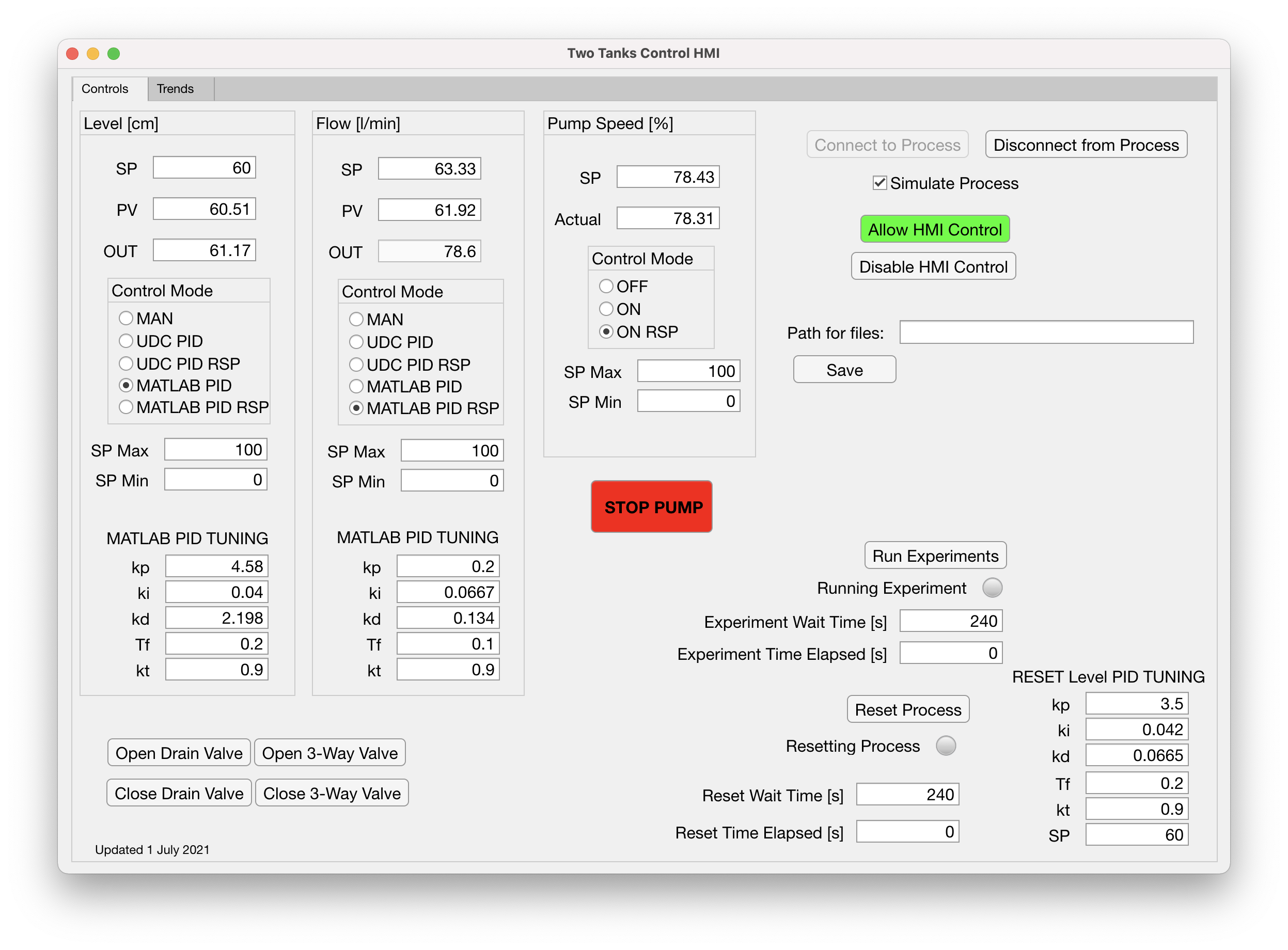}\label{fig:HMI_front}}
    \hfill\\
    \sidesubfloat[]{%
        \includegraphics[width=\linewidth, height=0.45\textheight, keepaspectratio]{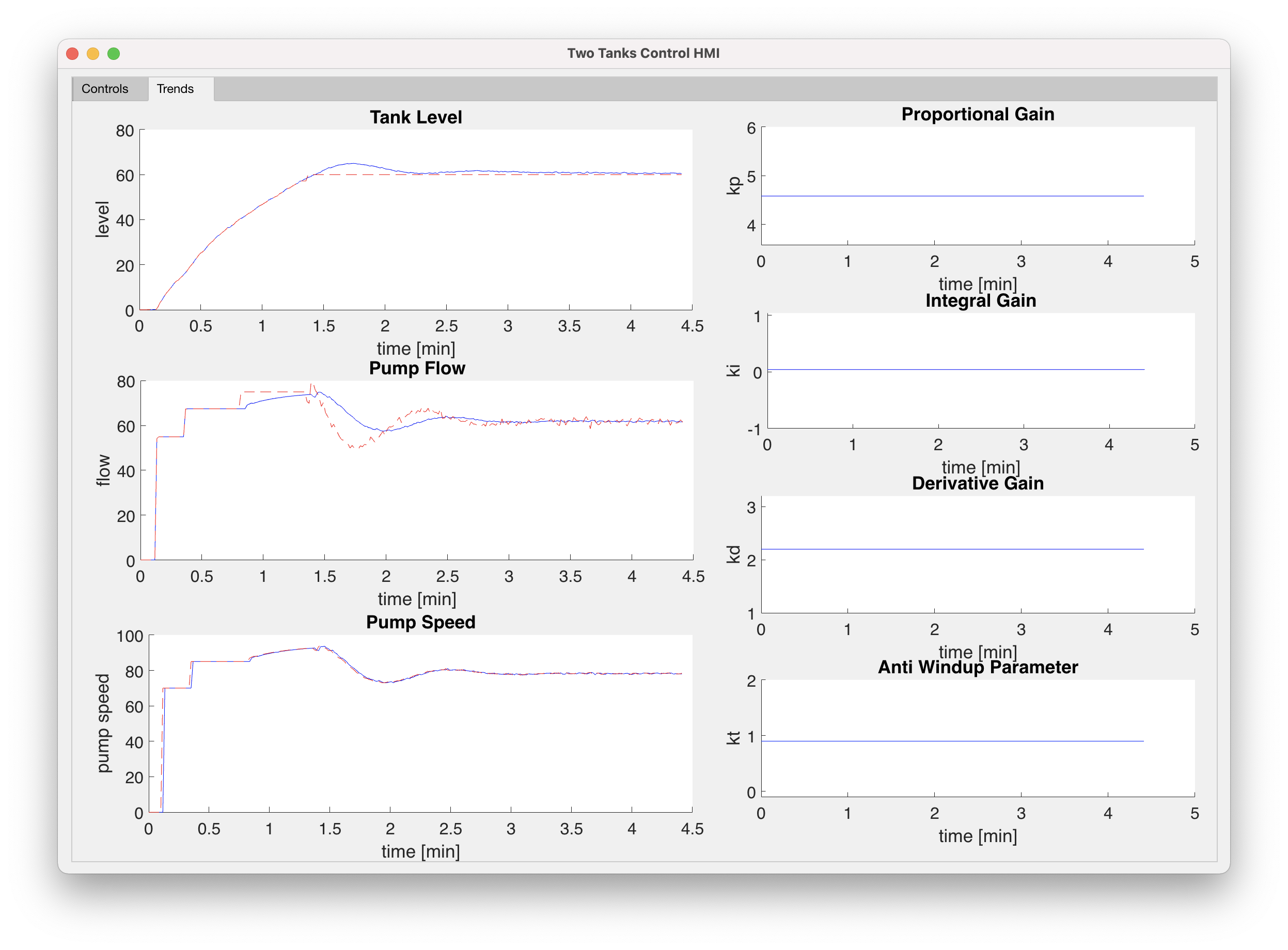}\label{fig:HMI_back}}
\end{center}
\end{minipage}
\caption{(a) The HMI control screen for the two-tank system. (b) Real-time plots show tracking performance and PID parameters; dashed lines are setpoints and solid lines are measured process variables.}
\label{fig:HMI}
\end{figure}
We developed a human-machine interface (HMI) in Matlab App Designer to interact with the tank system.
The interface is shown in \cref{fig:HMI_front}.
From the HMI, we can adjust the PID parameters for controlling the pump speed, flow rate, and level of the top tank.
More precisely, the PID controllers for the pump and flow produce setpoints which the physical pump must attain.
This is illustrated in \cref{fig:HMI_back}.
As a Matlab application, it is easy to build in a lot of additional functionality into the HMI. Although we refer to it as the HMI, it also includes implementations of PID controllers, automated and on-demand saving of data to csv files, and logic and algorithms to supervise and run process experiments. The HMI also includes an embedded simulation of the system based on the dynamics given in \cref{subsec:dynamics}.
This setup allows easy switching between simulation and reality, providing a unified interface for conceptual development and laboratory experiments.
We note that the simulator was \emph{not} used to pre-train the RL agent.\par

\subsection{Software and hardware for training RL agents}

\emph{(Software)\quad } Our starting point for developing a RL agent was the open source repository Spinning Up \citep{achiam2018SpinningDeep}.
In principle, deep RL libraries like Spinning Up can help streamline the process of testing new algorithms on novel simulated environments through the Open AI Gym \citep{brockman2016OpenAIGym}. However, our project called for a number of fundamental changes. Most importantly, the RL agent does \emph{not} directly actuate the system (as in Gym environments). Rather, it is one element of a modular design comprisingthe RL code, the HMI, and the instrumentation of the two-tank system.
\par
 In light of this, it is worth noting that there are two PID representations of \cref{eq:PIDlevel} at play: The actor network in the RL code, and the PID controller acting on the system through a hardware programmable logic controller (PLC). The actor network representation is implemented in PyTorch and used for the purposes of training. This setup requires inter-module communication to transmit the results of training to the PLC for implementation.


In order to accommodate this distinction, we made the following changes to the Spinning Up implementation of the TD3 algorithm:
\begin{itemize}[leftmargin=*]
    \item The standard actor network in Spinning Up is a feedforward neural network. A PID is a linear function followed by the identity activation function, given by \cref{eq:SS2}. We provided a custom initialization of the PID parameters (rather than the typical random initialization) and constrained the parameters to be positive through the use of a smooth and invertible approximation of ReLU ($x \mapsto \text{max}\{0,x\}$) called Softplus ($x \mapsto \ln(1 + \exp(x))$).
    For example, if one wishes to initialize the proportional gain as $k_p = 4.0$, then the corresponding weight in the RL code would be initialized as $\theta_{k_p} = \ln(\exp(4.0) - 1)$. A similar strategy constrains the anti-windup term to the interval $(0,1)$, by using Sigmoid ($x \mapsto 1/(1 + \exp(-x))$) instead of Softplus.
    Therefore, instead of taking $\theta = [k_p\ k_i\ k_d\ k_{\tau}]$ as in \cref{subsec:PIDRL}, the vector of inverted PID parameters is used as the trainable weights $\theta = [\theta_{k_p}\ \theta_{k_i}\ \theta_{k_d}\ \theta_{k_{\tau}}]$.
    \item We removed all linkages to a Gym-style environment. Instead, new data from the tank is processed in batches. Our HMI records process data and periodically saves it to a directory accessible to the RL agent. The RL code processes the new data then updates its internal representation of the PID controller according to \cref{alg:ActorCriticProposed}. More specifically, the RL code is responsible for reading these data files and constructing its replay buffer consisting of state transition tuples $(s, a, s', r)$. Once the weights $\theta$ are updated, they are translated back to ``PID form'' through the Softplus or Sigmoid functions as described above, then saved and read by the HMI.
    \item As a consequence of the above changes, we also removed all parameters in the code that would characterize an episode in the Gym environment. These include the number of time steps per episode, the number of samples to collect before training begins, and how many time steps pass between parameter (both actor and critic) updates. The effect of these parameters is still present, as the HMI includes these specifications; the main difference is that the RL code is only used to process new data. We can easily change design parameters in the HMI online, while the RL code is running in the background.
\end{itemize}

\emph{(Instrumentation and System Integration)\quad } The key field measurement devices and actuators included in the lab apparatus are a guided wave radar level measurement device to measure the level in the upper tank; a differential pressure flow measurement device to measure the flow of water into the upper tank; and a variable speed pump to pump water from the lower tank, through the flow measurement device, and into the upper tank. These are wired into an HC900 process and logic controller which allows communication with Matlab, via Modbus/TCP, to support a HMI and controls. Two UDC2500 loop controllers are also connected to the HC900 so that these elements may be used optionally for flow and/or level control. Logic programmed in the HC900 is used to switch between flow and level control by Matlab PID implementions, by the UDC2500 controllers, or by manual pump speed adjustments.

\emph{(Computing)\quad } For our purposes, it was sufficient to run the HMI and train the RL agent on a desktop computer in the lab. The PC used in the lab has an Intel Xeon CPU which runs at 3.5 GHz, and 8 GB of RAM. (This is a good but unremarkable PC.) Due to our modular setup, it would also be possible to train the RL agent through cloud services.

\section{Experiment design and case studies}

\subsection{Experiments for learning}

An episode is characterized by a step change in the process; a timer specifies how much time will elapse between step changes. To ensure safe operation even when unsupervised, the HMI has access to a set of PID parameters that are known to stabilize the system. These may be poorly tuned parameters, such as ones used before the tuning procedure. The HMI is able to switch to these parameters if the tracking provided by the RL-tuned parameters is too poor, guaranteeing that the next step change starts from steady state. Therefore, the timer parameter may be set online based on closed-loop experiments, serving as an upper bound for when to trigger the ``safe'' PID parameters unless the system is brought to steady-state beforehand. For simplicity and consistency in our experiments, we keep the timer fixed during experiments.
\par
Measured data are stored, processed, then used to train the RL agent during operations. The RL agent generates new PID parameters, which are then loaded into the HMI to update the Matlab PID control implementations. The outputs from these PID controllers are sent to the HC900 process and logic controller which passes them to the appropriate field devices. This process of sending new data to the RL agent and updating the PID parameters can be done during an experiment or set to occur at the end.\par
We train using a pre-defined set of setpoints; an ``episode cycle'' refers to one pass over all of these setpoints. In the context of training, an ``experiment'' refers to all the episodes leading to the final tuning parameters. Later on we refer to an ``evaluation experiment'' or ``robustness experiment'' in the context of some final set of tuning parameters that we wish to evaluate in various conditions. \par

\subsection{Reward function selection}
\label{subsec:reward}

A crucial component of any RL agent is the reward function. This function must accurately capture the goals of the system designer.
Therefore, we consider costs (negative rewards) that depend only on the tracking error $e_t$ and the change in control variable $\Delta u_t = u_t - u_{t-1}$, in the form
\begin{equation}
    l(s_t, u_t) = \abs{e_t}^p + \lambda \abs{\Delta u_t}^q,
\label{eq:rewardFunc1}
\end{equation}
where $p,q$ are fixed integers (namely, $1$ or $2$) and $\lambda \geq 0$ is a fixed penalty term. For a system with multiple inputs and outputs, one can generalize \cref{eq:rewardFunc1} by using the $\ell_1$ or squared $\ell_2$ norms. Although the cost given by \cref{eq:rewardFunc1} is commonly used in the case of $p=q=2$, such as in applications of model predictive control, the user is left with the tuning parameter $\lambda$. A reasonable initial choice is $\lambda = 1/\abs{\Delta u}_\text{max}$, where $\abs{\Delta u}_\text{max}$ is an approximation of the maximum absolute value of $\Delta u$ observed within the operating region. This is simply a normalization step: one can rewrite the penalty term as $\lambda \frac{\abs{\Delta u_t}^q}{\abs{\Delta u}_\text{max}}$ and adjust $\lambda$ based on relative weight given to the tracking error term of \cref{eq:rewardFunc1}. For our purposes, we use the penalty term $0.1 (\Delta u_t)^2$.\par
The above reward function structure is flexible in terms of the range of behaviors it can incentivize. For example, setting $\lambda=0$ selects the absolute or squared error reward function. The squared error places a significant amount of weight on states with large errors; in comparison, the absolute error puts more emphasis on small errors. Therefore, it is reasonable to prefer the absolute error for a slower response but with better attenuation of overshoot and oscillations. Conversely, a fast response may be desirable. One of the appeals of RL is the flexibility in the choice of the reward function. Therefore, we also test a cost function that has the desirable components of both the absolute and squared errors:
\begin{equation}
    l(s_t, u_t) =
    \begin{cases}
        \abs{e_t} &\quad \text{if}\ \abs{e_t} < 1\\
        \frac{1}{2}(e_t^2 + 1) &\quad \text{otherwise.}
    \end{cases}
\label{eq:hybridReward}
\end{equation}
\cref{eq:hybridReward} is the absolute value function around the origin and smoothly transitions to a parabola. From now on, we use $l$ exclusively for \cref{eq:hybridReward}, and refer to it as the ``hybrid cost'' (or reward). One may set $r=-l$ or use a reward based on \cref{eq:hybridReward} with an input penalty.

\subsection{Tuning subject to input constraints}
\label{subsec:inputConstraint}

One difficulty in applying RL algorithms to tune a PID controller is the inherent presence of input constraints in a physical system. We follow the methods proposed by \citet{hausknecht2016DeepReinforcement}: The actor update scheme in \cref{eq:actorgradient} is modified to steer its actions to within a pre-defined range $[u_\text{min} , u_\text{max}]$. Formally, since
\begin{equation}
\nabla_{\theta} Q_{\phi_i}(s, u)|_{u = \mu_{\theta}(s)} = \nabla_{\theta}\mu_{\theta}(s) \nabla_u Q_{\phi_i}(s, u)|_{u = \mu_{\theta}(s)} 
\label{eq:critic_grad}
\end{equation}
by the chain rule, the components of $\nabla_u Q_{\phi_i}(s, u)|_{u=\mu_{\theta}(s)}$ are scaled as follows:
\begin{align}
\frac{\partial Q_{\phi_i}}{\partial u}(s,u) \leftarrow
\frac{\partial Q_{\phi_i}}{\partial u}(s,u) \cdot
\begin{cases}
\dfrac{u_\text{max}-u}{u_\text{max}-u_\text{min}} & \text{if}\quad \frac{\partial Q_{\phi_i}}{\partial u}(s,u) > 0 \\
\dfrac{u-u_\text{min}}{u_\text{max}-u_\text{min}}              & \text{otherwise}.
\end{cases}
\label{eq:Gradients}
\end{align}
Without making the RL agent ``aware'' of the input constraints it may continue to propose infeasible actions. Intuitively, \cref{eq:Gradients} reverses the direction of the parameter update if the critic ``reinforces'' an infeasible action toward the optimum. This update scheme also puts less weight on a parameter update when an action is close to the constraints.\par
Another option for updating the actor parameters subject to constraints is to use an output activation on the actor such that the actions are automatically forced inside the range $[u_\text{min} , u_\text{max}]$ \citep{hausknecht2016DeepReinforcement}. Options include using the saturation function or a smooth approximation such as tanh. However, this approach diminishes the gradient in \cref{eq:critic_grad} at the constraints, essentially ignoring the value of such state-action pairs. We emphasize that the PID controller on a physical system will still obey the constraints $[u_\text{min} , u_\text{max}]$, whether by physical limitations or by modeling them directly with the saturation function; the update scheme in \cref{eq:Gradients} is solely for the purpose of updating the actor and any constraint violations performed in the update scheme are not reflected on the physical system.

\subsection{Standard tuning methods}
\label{subsec:tuningmethods}

To provide an overall evaluation of our RL algorithm, we compare our results to baseline tuning methods in the context of the criteria put forth in \cref{sec:scorecard}. In particular, we compare the RL results to various tuning parameters given by the Skogestad IMC (SIMC) tuning method \citep{skogestad2003SimpleAnalytic} and Honeywell's Accutune~III algorithm \citep{honeywell2007UDC2500Universal, berner2018ExperimentalComparison}. SIMC provides PI parameters based on a first-order plus dead time (FOPDT) model of the plant; the closed loop time constant $T_c$ is the only tuning parameter for this method. Accutune~III is a relay autotuning algorithm; its user inputs are the input range for the relay signal and a switch to select either ``fast'' or ``slow'' tuning. Note that these methods do not share the same underlying objective of RL, so the purpose of including them is simply for baseline data of what reasonable performance or robustness look like in our setting. Moreover, we can evaluate the three methods (RL, SIMC, Accutune) through the lens of our ``scorecard'' items in \cref{sec:scorecard}. \par

\section{Lab results}
\label{sec:results}

We report our experimental results in two sections. This section reports some key results and high-level conclusions. \ref{app:experiments} provides all the data supporting our findings. \cref{table:table_hyperparameters} lists the hyperparameters used in this work. We emphasize that all the experiments presented here are performed directly on the physical two-tank system without prior pre-training, for example, in simulation or with offline datasets.\par

\subsection{Experimental setup}

We run three different RL experiments, all using this reward function:
\begin{equation}
-r(s_t, u_t) = \abs{e_t} + 0.1 \Delta u_t^2\label{eq:baselineReward2}.
\end{equation}
They differ based on the initial PID tuning parameters for training as well as the operating conditions. The first two experiments are ``unconstrained experiments''; more precisely, we run \cref{alg:ActorCriticProposed} in an operating region where the agent is unlikely to hit the physical constraints of the pump. Our algorithm is designed to improve the initial tuning parameters; therefore, we test the algorithm from different starting parameters. The initial tuning parameters are given according to the following dependence on $k_p$: 
\begin{equation}
k_i = k_p / 60,\ k_d = 0.01k_p,
\label{eq:initPID}
\end{equation}
where the first experiment sets $k_p = 4.0$ and for the second $k_p = 2.0$. Since the RL agent operates in closed loop, the input constraints may not always be avoidable, which is why they must be accounted for during training. We next run a ``constrained experiment" with $k_p = 4.0$. That is, we run \cref{alg:ActorCriticProposed} with a strategy for dealing with input constraints, as described in \cref{subsec:inputConstraint}. This experiment switches between the setpoints $60$~cm and $65$~cm as well $60$~cm and $63$~cm; this configuration gives the agent training data for setpoint tracking both with and without input constraints.\par

\subsection{Evaluation procedure}

For each set of tuning parameters (five in total), we cycle through each setpoint for four minutes each. The setpoint sequence is $60$~cm, $65$~cm, $60$~cm, $63$~cm. Input constraints are inactive for the evaluation. The reason for this, even for the results from the constrained experiment, is to evaluate the quality of the PID parameters across all methods.\par

For each of these evaluation experiments, we calculate the average normalized integral absolute error (IAE), integral squared error (ISE), total variation (TV), total variation in the input variable ($\text{TV}_u$), percent overshoot (\% OS), and settling time (ST). \emph{Normalized} performance means the IAE, ISE, TV use the error signal divided by the change in the setpoint for that step change in the calculation;  $\text{TV}_u$ is calculated by dividing by the initial $\Delta u$ value instead.\par

To perform the SIMC tuning method we first derive a FOPDT model of the flow setpoint to level dynamics of our system. Based on data from stepping the flow setpoint up and down to steady state, we obtain the model $G(s) = \frac{3.44}{301.19 s + 1} e^{-9.21 s}$. We consider the values $T_c = 9.21, 15, 20, 25, 30$. The first value of $T_c$ comes from the ``default'' configuration of setting $T_c$ equal to the process time delay; however, since $G$ is nearly an integrating processes, this setting may not be desirable. In this section, we only use $T_c = 20$; for completeness, the other values are evaluated in \ref{app:experiments}. Finally, we use a UDC2500 to run the Accutune III procedure and the control the flow for the ensuing evaluation steps.\par 

\subsection{Summary of results}

\begin{table}[!tbh]
\begin{center}
\resizebox{\linewidth}{!}{%
\begin{tabular}{Nlllllll}
\toprule
\multicolumn{0}{l}{} &    IAE &    ISE &    TV & TV$_u$ & $\%$ OS &      ST & $\text{M}_s$ \\
\midrule
RL\label{row:nominal_RL}   &  $39.34\pm3.44$ &  $27.56\pm2.25$ &  $1.49\pm0.08$ &   \phantom{00}$9.37\pm0.42$ &    $6.39\pm1.21$ &   $127.00\pm38.80$ & $1.32\pm0.06$ \\
SIMC\label{row:nominal_SIMC20} &  47.74 &  32.31 &  1.50 &   \phantom{00}9.14 &    7.81 &  215.25 & 1.25\\
Accutune\label{row:nominal_AccuSlow} &  54.06 &  34.06 &  1.75 &  583.78 &    9.14 &  159.75 & 1.41\\
\bottomrule
\end{tabular}
}

\end{center}
\caption{The nominal performance of RL experiments and baseline tuning methods. Each cell shows the average normalized performance over the sequence of step changes.} \label{table:table_subsetExperiments_nominal}
\end{table}

\cref{table:table_subsetExperiments_nominal} shows the nominal performance of the final tuning parameters across the RL, SIMC, and Accutune evaluation experiments. We also include the maximum sensitivity $\text{M}_s = \max_{0 \leq \omega < \infty} \left( 1 + C(i \omega) G(i \omega) \right)^{-1}$ based on the model $G$ used for SIMC tuning. The RL row summarizes the performance the three aforementioned experiments. We report the average of each of these statistics across the three experiments, plus or minus the standard deviation. All three experiments perform better than Accutune III across all metrics, with the one exception of ST for the constrained RL experiment. RL and SIMC achieved similar TV, $\text{TV}_u$ and OS, with RL performing better in terms of IAE, ISE, and ST. Ultimately, the result of the RL tuning across these different operating conditions is smooth and efficient tracking. Moreover, as we will see next, each RL result was obtained in around 40 minutes of operation.\par

\begin{figure}[!phtb]
\begin{minipage}[]{\linewidth}
\begin{center}
    \sidesubfloat[]{%
      \includegraphics[width=.40\linewidth, keepaspectratio]{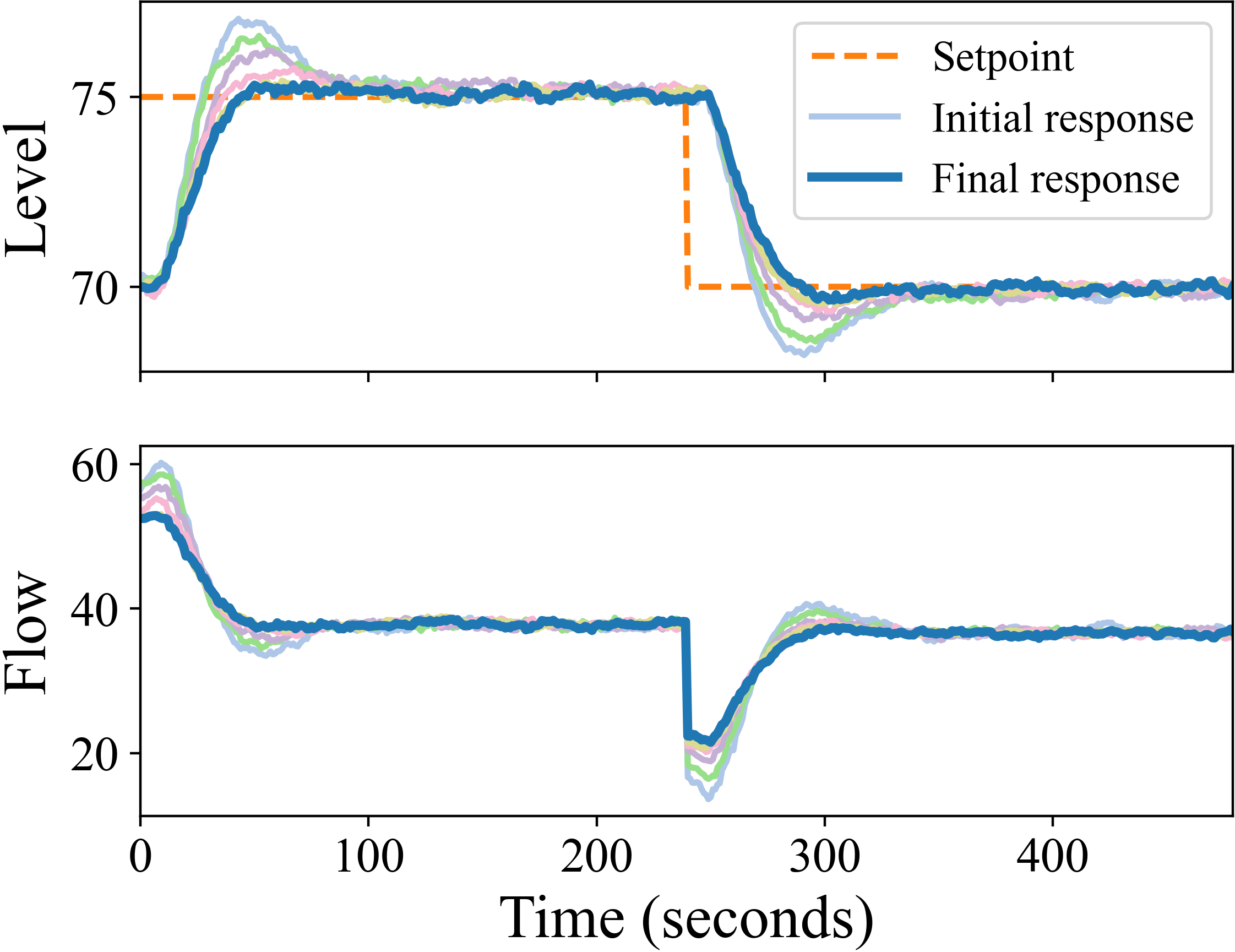}\label{fig:experiment_td3_2021_04_07_1208}}\hfill%
    \sidesubfloat[]{%
      \includegraphics[width=.40\linewidth, keepaspectratio]{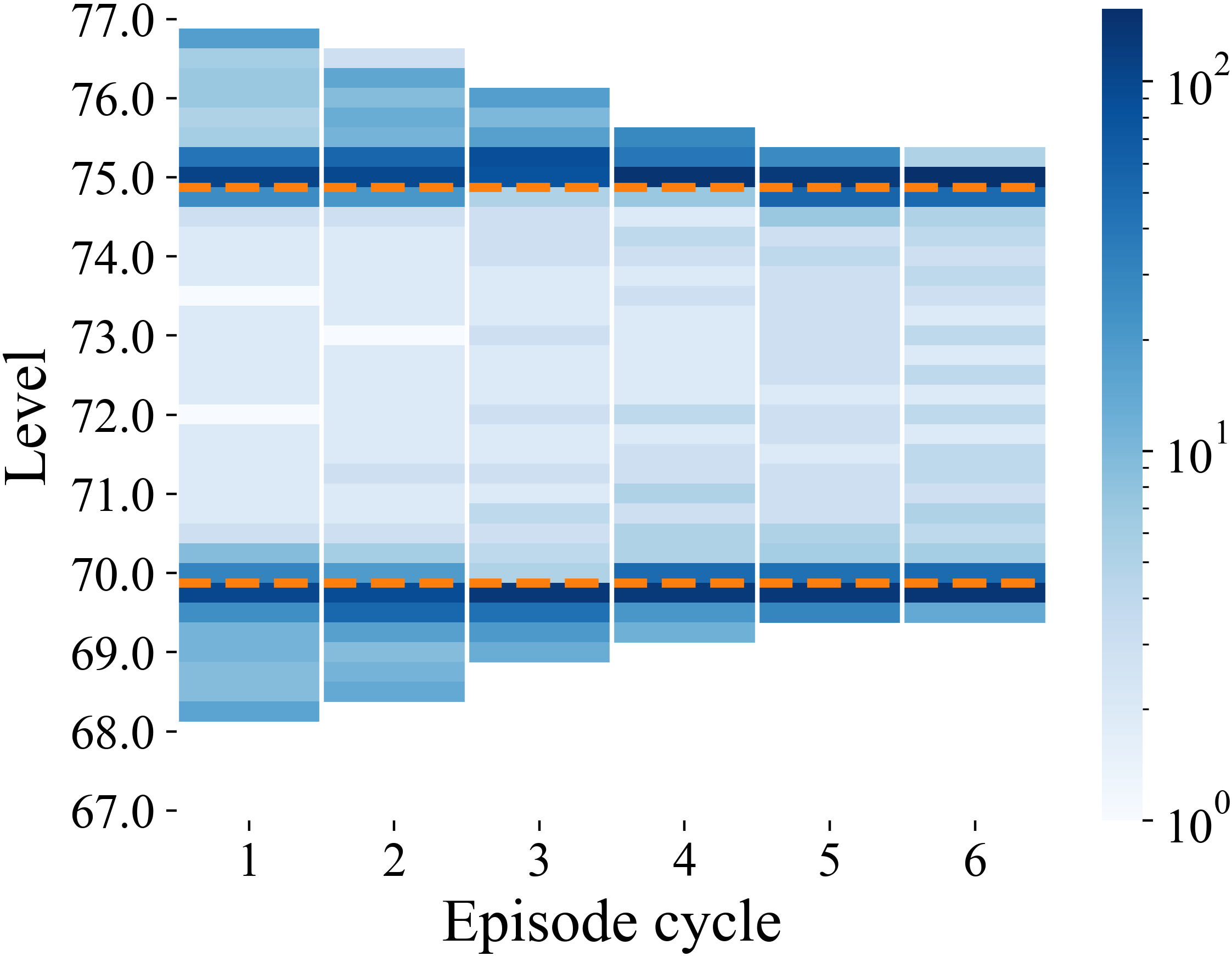}\label{fig:heatmap_td3_2021_04_07_1208}}\hfill\\
    \sidesubfloat[]{%
    \includegraphics[width=.40\linewidth, keepaspectratio]{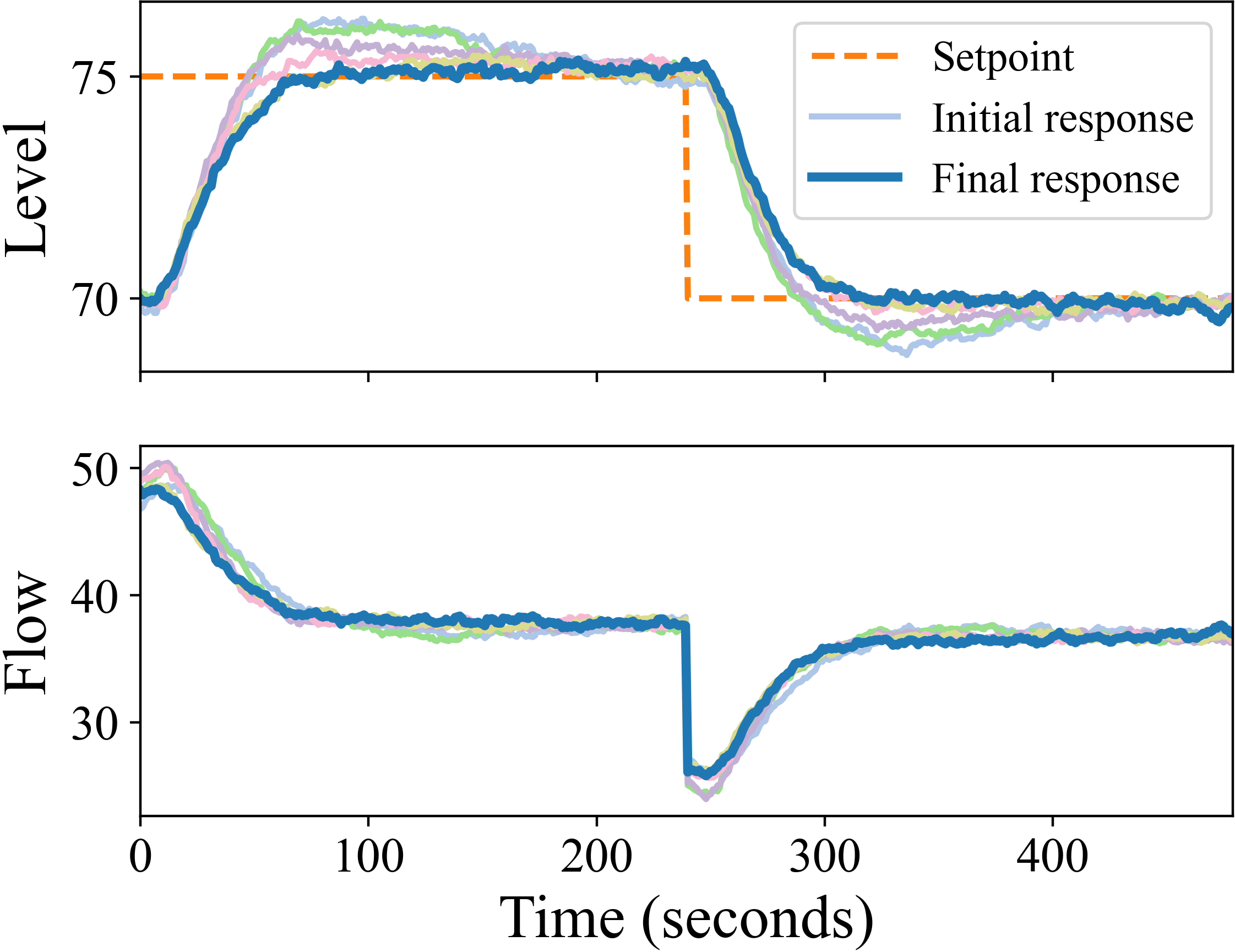}\label{fig:experiment_td3_2021_04_07_1357}}\hfill
    \sidesubfloat[]{%
    \includegraphics[width=.40\linewidth, keepaspectratio]{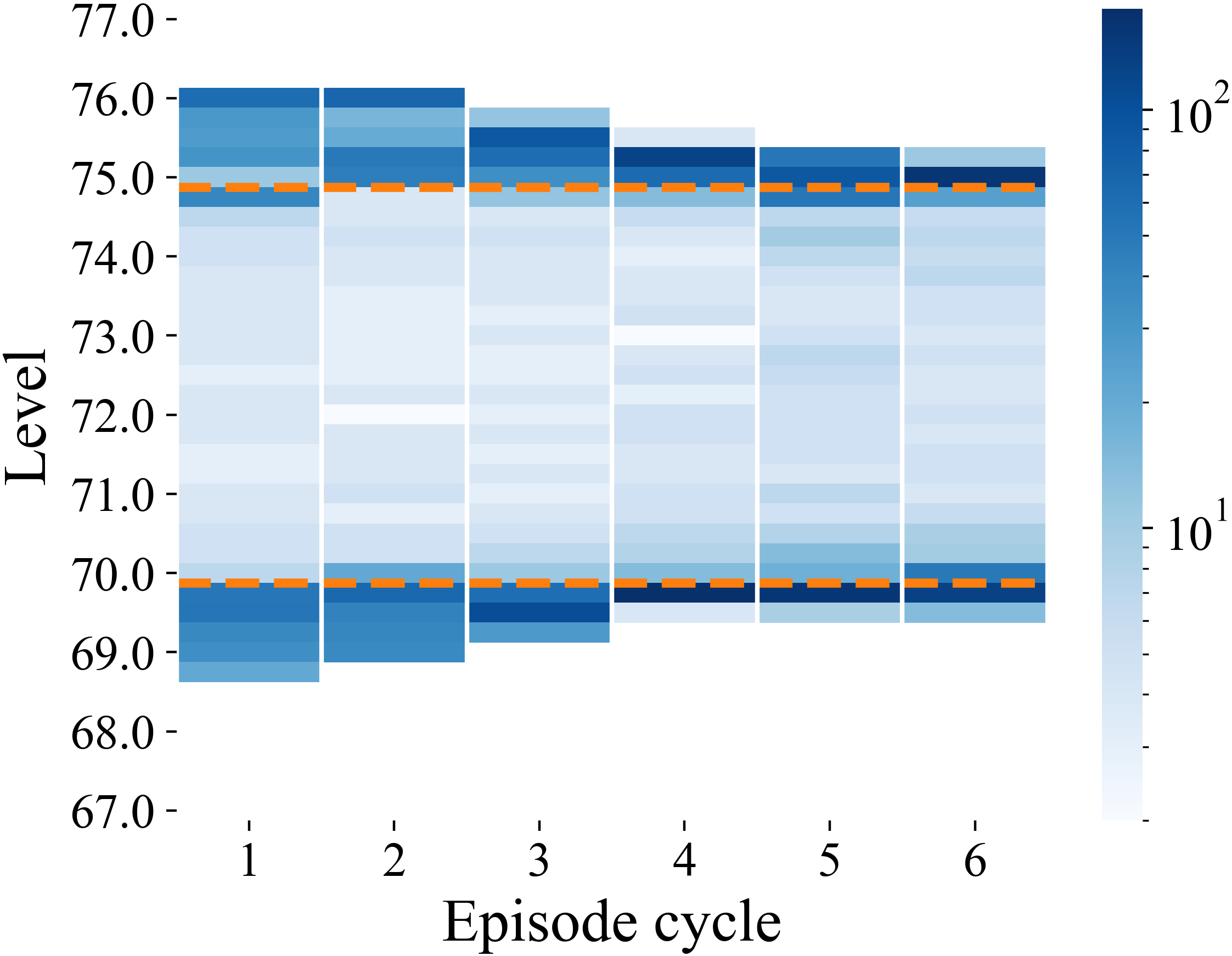}\label{fig:heatmap_td3_2021_04_07_1357}}\hfill
\caption{Performance evolution under two different reward functions: 
(a) Tracking in experiment for \rowref{RL}{row:robustness_td3_2021_04_07_1208}; 
(b) Tracking heatmap for \rowref{RL}{row:robustness_td3_2021_04_07_1208}: 
$x$-axis is progression of episode cycles ($2$ step changes each) for training and the $y$-axis is the level; 
(c) Tracking in experiment for \rowref{RL}{row:robustness_td3_2021_04_07_1357}; 
(d) Tracking heatmap for \rowref{RL}{row:robustness_td3_2021_04_07_1357}.}
\label{fig:experiment_heatmap}
\end{center}
\end{minipage}
\end{figure}

\cref{fig:experiment_heatmap} shows the evolution of the training process for the two unconstrained experiments. We show two visualizations for each experiment ($k_p = 4.0$ and $k_p = 2.0$, respectively). Time-series plots are given in \cref{fig:experiment_td3_2021_04_07_1208} and \cref{fig:experiment_td3_2021_04_07_1357}. The initial step performances are characterized by a fast rise with significant overshoot (roughly 40\%) or a slow settling time (roughly 3.5 minutes). We see that in all the experiments the performance uniformly plateaus at around 10 episodes. Each episode is roughly four minutes. Even though the experiments run for varying lengths of time, all of them reach their peak performance in less than 40 minutes of operation.\par 
\cref{fig:heatmap_td3_2021_04_07_1208} and \cref{fig:heatmap_td3_2021_04_07_1357} are respective heatmaps of the same output data. The time-series and heatmaps are shown side-by-side to convey the intuition for the heatmap, which is more heavily used in \ref{app:experiments} because it is a compact way of showing many different experiments together. The darker shades mean the process variable spent more time in that region of the $y$-axis than lighter regions. For example, the heatmap captures overshoot with the presence of shaded regions above the setpoint, and conveys settling time based on the distribution of shading around the setpoint. We see as training progresses (that is, as the number of episodes increases) the darker shading is more concentrated around the setpoints and the overshoot decreases. We also see a slightly longer rise time based on how dark the region is \emph{between} setpoints.\par 
\begin{figure}[!htb]
\begin{minipage}[]{\linewidth}
\begin{center}
    \sidesubfloat[]{%
      \includegraphics[width=0.95\linewidth, height=0.40\textheight, keepaspectratio]{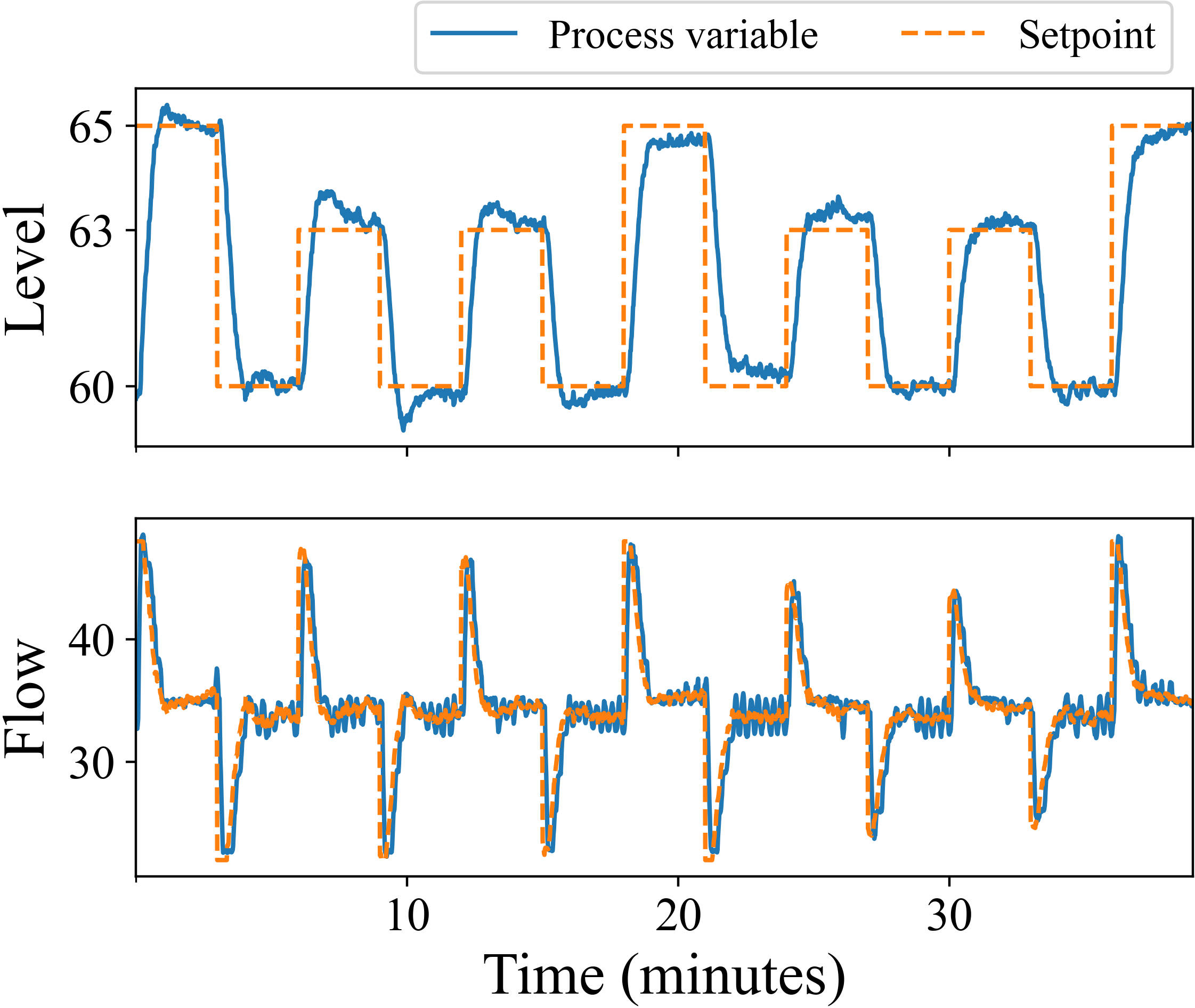}\label{fig:timeseries_td3_2021_04_28_1451}}\hfill\\
    \sidesubfloat[]{%
      \includegraphics[width=.70\linewidth, height=0.40\textheight, keepaspectratio]{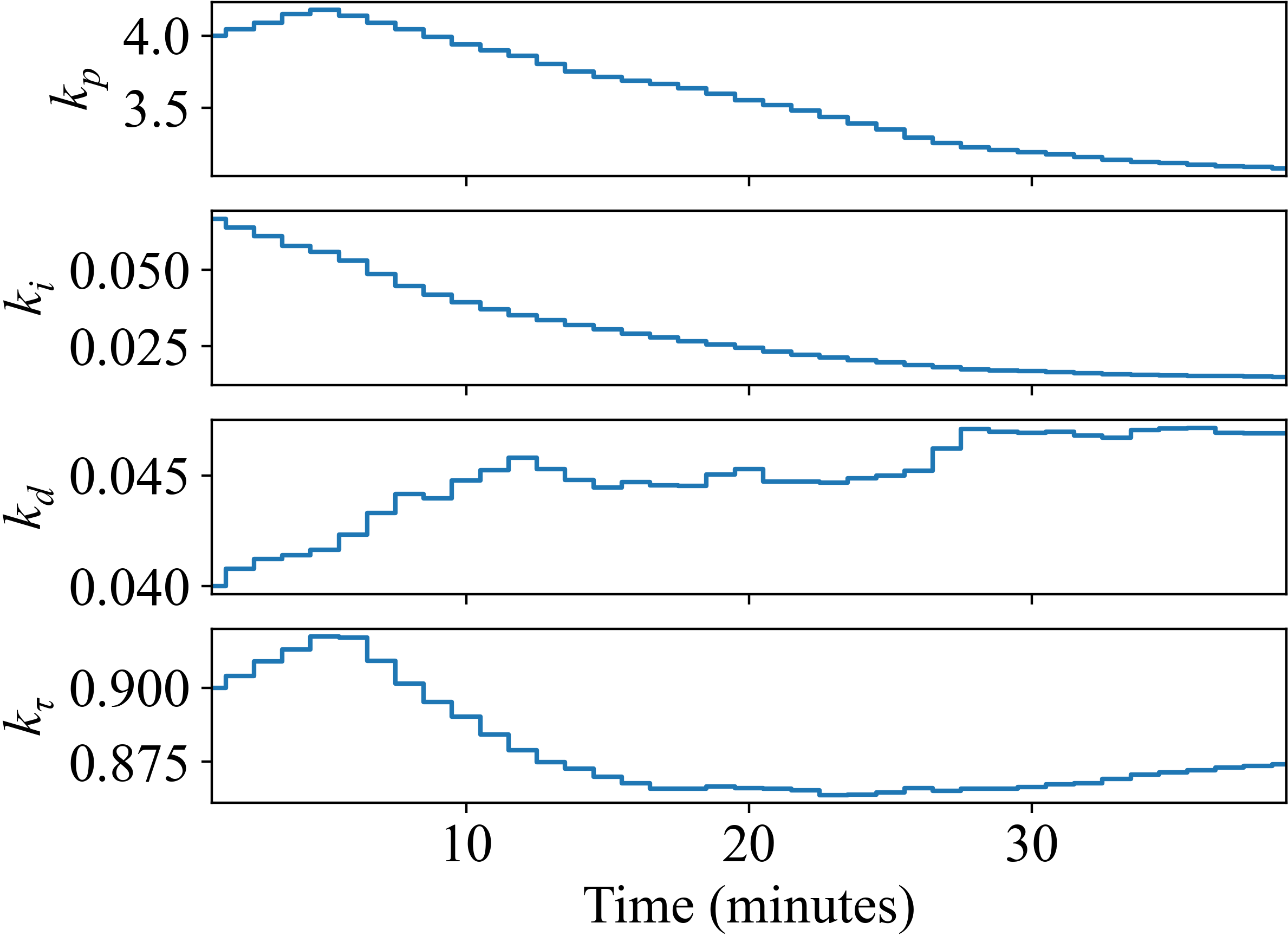}\label{fig:params_td3_2021_04_28_1451}}\hfill\\
\caption{The progression from initial to final performance of \rowref{RL}{row:robustness_td3_2021_04_28_1451}. (a) A time-series of the entire training procedure; (b) A temporally-aligned plot of the tuning parameters during training.}
\label{fig:experiment_constrained}
\end{center}
\end{minipage}
\end{figure}

The two RL experiments corresponding to $k_p = 4.0$, one unconstrained and one constrained, achieved similar evaluation performance with the exception of ST, where the unconstrained experiment achieved roughly 40\% better settling time. However, the latter was trained subject to more realistic conditions. Further, the qualitative end result of the constrained experiment, as shown in \cref{fig:experiment_constrained}, is still excellent. Note that this time-series is with the input constraints in place and is also temporally-aligned with the tuning parameters during training.\par 
Despite the differences among RL experiments, all but two of them, which we show in \ref{app:experiments}, performed better than SIMC in terms of IAE and ISE. Two instances of SIMC showed an advantage over RL in terms of TV, $\text{TV}_u$, and $\% \text{OS}$, but at a significant cost in IAE and ISE. \rowref{Accutune}{row:robustness_AccuSlow} performs reasonably well in terms of TV, $\% \text{OS}$, and ST, but achieves, by far, the highest values of IAE, ISE, and $\text{TV}_u$ across the RL and SIMC experiments. Finally, the absolute value based reward is good for ``smooth'' tracking and overall performs well across the metrics in \cref{table:table_subsetExperiments_nominal}. In \ref{app:experiments}, rewards based on the squared error or hybrid function may be better for ``fast" tracking but suffer in terms of TV, $\text{TV}_u$, and $\% \text{OS}$ as well as in experimental robustness. We consider the experiments based on the reward in \cref{eq:baselineReward2} to be the overall best option.\par

\subsection{Overall evaluation}
\label{subsec:evaluation}

\begin{table*}[!tb]
\begin{center}
\begin{tabularx}{0.80\linewidth}{lYYYYYYY} 
\multicolumn{1}{c}{} & \multicolumn{1}{c}{\rot{Nominal performance}} & \multicolumn{1}{c}{\rot{Robustness}} & \multicolumn{1}{c}{\rot{Time required for tuning}} & \multicolumn{1}{c}{\rot{Disturbance to process}} & \multicolumn{1}{c}{\rot{Hardware requirements}} & \multicolumn{1}{c}{\rot{Ease of use}} & \multicolumn{1}{c}{\rot{Input constraints}}\\
\midrule
SIMC & -- & -- & -- & \xmark & \cmark & -- & \xmark \\
Accutune III & \xmark & \xmark & \cmark & \cmark & \cmark & \cmark & \xmark \\
Deep RL & \cmark & \cmark & \xmark & \cmark & -- & \cmark  & \cmark \\
\bottomrule
\end{tabularx}

\end{center}
\caption{A qualitative evaluation of deep RL, SIMC, and Accutune III. We use \cmark\ to indicate ``excellent''; \xmark\ to indicate ``poor''; and\ --\ to indicate ``neutral''}
\label{table:scorecard}
\end{table*}

As mentioned in \cref{sec:scorecard}, performance and robustness are not the only factors to consider when evaluating a tuning method. \cref{table:scorecard} provides a summary of our qualitative assessment of deep RL with SIMC and Accutune~III included as points of reference. Unsurprisingly, deep RL achieved excellent nominal performance across different initializations, reward functions, and subject to input constraints. Moreover, the tuning of these results is robust to changes in the two-tank system (see \ref{app:experiments}) . Although these result were achieved in a reasonable amount of time ($30-50$ minutes), commercial auto-tuners such as Accutune~III provide tuning parameters in roughly $10$ minutes (excluding additional time to evaluate the performance). SIMC recommends setting $T_c = \theta$, but manipulating $T_c$ from there requires additional experiments. The next thing to note is that SIMC and Accutune~III are open loop tuning methods, whereas deep RL operates in closed loop; therefore, despite longer training time, we consider the disturbance to the process to be more practical than SIMC or Accutune~III, \emph{if} one can provide an acceptable range of setpoints. Accutune~III is also desirable in this regard because it switches the relay signal based on deviations of the process variable from the current setpoint. An advantage of SIMC and Accutune~III is that they are simple enough such that no specialized hardware is required. For a single control loop, we were able to use a standard desktop for training the RL agent. Finally, taken in their final form, both deep RL and Accutune~III can be simple to use: in the case of RL, a user may input some rudimentary information, such as acceptable setpoints, while Accutune~III relies on an interval of admissible inputs.

\section{Discussion and conclusion}
\label{sec:conclusion}

To conclude this study, we contrast our findings with some common themes that circulate in the deep RL literature and highlight promising areas for future work.

The primary concerns at the prospect of applying deep RL in the process industries pertain to stability, interpretability, sample efficiency, and practicality \citep{nian2020ReviewReinforcement, shin2019ReinforcementLearning}. In some ways these are complementary concepts: In the broad landscape of RL, the policy is often represented by a DNN, which makes it difficult to rigorously explain its behavior. This contributes to the difficulty surrounding sample efficiency, interpretability, and stability due to the nonlinear structure of a DNN operating in a closed-loop system. From a practical implementation perspective, it is of course possible to deploy these policies in an industrial control system. However, there already exists extensive investment and research into deploying control architectures such as MPC and PID. Given their prevalence and track record, the most promising starting point for RL applications in the process industries is through some hybrid approach.\par
In this work, we have directly parameterized the policy as a PID controller and configured it by solving the RL objective using a model-free deep RL algorithm. Other approaches, mentioned in \cref{subsec:related}, train the RL policy to output PID parameters as ``actions'', but the policy itself is a DNN or derived from value-based approaches. Consequently, we have orders of magnitude fewer parameters to tune. Crucially, we also exploit the fact that a PID controller is standard in universal digital controllers, and therefore only need to send new parameters to the controller, rather than install new hardware in order to run the RL policy.\par
In our experimental results we are encouraged by the monotonic improvement in IAE and ISE, ultimately leading to a well-tuned PID controller within $30-50$ minutes. We attribute this to the small number of parameters in a PID controller and the fact that setpoint tracking is a ``primitive'' of its design; in other words, it does not need to ``learn'' how to track setpoints, rather improve upon its existing performance. Note that these results are obtained by training the critic network from scratch, that is, without any pre-training either from a simulation or historical process data. The fact that we could deploy the algorithm without adding any specialized hardware to the system, while operating in closed-loop, and performing the computation on a local desktop computer is encouraging. Ultimately, these are promising results on which to build.\par

\subsection{Opportunities in deep RL}
\label{subsec:opportunities}

Despite some promising results presented here, a looming question persists: How can one systematically configure the RL agent to a novel environment when it has failed to learn? The central problem at play is that of training RL agents on a case-by-case basis. Despite being dubbed ``model-free'', actor-critic aims to capture the system dynamics by directly modeling the value function, which is used to update the controller. More generalized, offline approaches are a promising avenue forward aim to achieve this over a range of similar systems or with historical datasets alone. Several frameworks have been proposed: \emph{offline RL} \citep{levine2020OfflineReinforcement} aims to train the RL agent using \emph{only} historical data from the system of interest. These methods are concerned with training a predictive model or value function that accounts for the uncertainty between the data samples and environment. On the other hand, \emph{meta-RL} \citep{mcclement2021MetaReinforcementLearning} trains an agent to learn from a distribution of similar dynamics and objectives; that is, the RL agent is not only trained to achieve optimal control, but also to learn an encoding of its environments, enabling it to generalize its policy to new systems. Latent representations of the system dynamics have been shown to be critical elements achieving real-world generalization from training in simulation in robotics applications \citep{lee2020LearningQuadrupedal}.\par  
The ability to train a generalized RL agent and utilize historical data from individual systems is a significant opportunity for the process industries. We believe the primary benefits of these approaches are twofold: Increased scalability of RL algorithms and safety of the training procedure. As mentioned earlier, an outstanding issue with RL algorithms is the ability to reliably choose hyperparameters in the event of poor training performance. Therefore, training offline is a safety precaution. Moreover, algorithms that can effectively distill information from a range of different systems into a single agent will increase the scalability of RL by decreasing the cost of calibrating the agent to novel environments.

\section*{Declaration of competing interest}
\label{sec:Declaration}
The authors declare that they have no known competing financial interests or personal relationships that could have appeared to influence the work reported in this paper.

\section*{Acknowledgements}
\label{sec:acknowledgements}
We gratefully acknowledge the financial support of the Natural Sciences and Engineering Research Council of Canada (NSERC) and Honeywell Connected Plant. We would like to thank Jude Abu Namous for helping us obtain lab results by running countless experiments at Honeywell in North Vancouver. We would also like to thank, from Honeywell, Shadi Radwan for designing and assembling the two-tank system, and Stephen Chu for his help with instrumentation and system integration.







\footnotesize
\bibliographystyle{elsarticle-num-names}
\bibliography{main.bib}
\normalsize







\appendix

\section{Further algorithmic and implementation details}
\label{app:details}

{
\begin{minipage}[]{\linewidth}
\linespread{1}
\footnotesize
\SetArgSty{textnormal}
\begin{algorithm}[H]
\caption{Deep Reinforcement Learning PID Controller}
\label{alg:ActorCriticProposed}
\DontPrintSemicolon
\setlength{\abovedisplayskip}{0pt}
\setlength{\belowdisplayskip}{0pt}
\setlength{\abovedisplayshortskip}{0pt}
\setlength{\belowdisplayshortskip}{0pt}
\SetKwInput{KwInitialize}{Initialize}
\SetKwInput{KwOutput}{Output}
\SetKwComment{Comment}{$\triangleright$\ }{}
\KwOutput{Optimal PID controller $\mu_{\theta}(\cdot)$}
\KwInitialize{Actor tuning parameters $\theta$, critic weights $\phi_1$, $\phi_2$, measurement dataset $\mathcal{D}_{\text{process}}$, replay memory $\mathcal{D}_{\text{RM}}$, step sizes $\alpha_a, \alpha_c$}
Set target parameters equal to actor/critic parameters $\tilde{\theta}\leftarrow \theta$ and $\tilde{\phi_1}\leftarrow \phi_1$, $\tilde{\phi_2}\leftarrow \phi_2$\;
\For{each episode}{
\Comment*[l]{In HMI:}
Set $\bar{y} \leftarrow$ setpoint\;
    \Repeat{next step change or system reaches steady state}{
    \Comment*[l]{In HMI:}
    Load current PID parameters $\theta$\;
    Observe state $s$ from environment\;
    Execute control action $u \leftarrow \mu_{\text{PLC}}(s)$ \Comment*[r]{$\mu_{\text{PLC}}$ is a PLC implementation of the actor ``network'' $\mu_{\theta}$.}
    Observe next state $s'$ from environment\;
    Store process data in $\mathcal{D}_{\text{process}}$\;
    \Comment*[l]{Execute in Python:}
        \If{it is time to update}{
    		Store transition tuples $(s, u, r, s')$ in $\mathcal{D}_{\text{RM}}$ \Comment*[r]{These structured tuples are derived from $\mathcal{D}_{\text{process}}$.}
            \For{each update step j}{
            \label{line:sample} Randomly sample a batch of transitions $\mathcal{B}$ from $\mathcal{D}$\;
            Compute target actions $a' = \sat(\mu_{\tilde{\theta}}(s') + \sat(\epsilon))$, $\epsilon \sim \mathcal{N}(0, \sigma^2)$\;
            Compute targets $q = r + \gamma \min_{i=1,2}{Q_{\tilde{\phi_i}}(s', a'})$\;
            Update critic weights as follows  for $i=1,2$:\;$\phi_i \leftarrow \phi_i - \alpha_c \nabla_{\phi_i} \left( \frac{1}{\abs{\mathcal{B}}} \sum_{(s, u, r, s') \in \mathcal{B}}(q-Q_{\phi_i}(s,a))^2 \right)$\label{line:critic_update}\;
                \If{$j \mod$ \texttt{policy\_{delay}} $ = 0$}{
                Update policy weights \Comment*[r]{Optional: Use \cref{eq:Gradients}.} $\theta \leftarrow \theta + \alpha \nabla_{\theta} \left( \frac{1}{\abs{\mathcal{B}}} \sum_{s \in \mathcal{B}} Q_{\phi_1}(s, \mu_{\theta}(s)) \right)$\label{line:actor_update}\;
                Update target weights
                \begin{algomathdisplay}
                \begin{aligned}
                \tilde{\phi_i} &\leftarrow \rho \tilde{\phi_i} + (1-\rho) \phi_i \quad \text{for } i=1,2\\
                \tilde{\theta} &\leftarrow \rho \tilde{\theta} + (1-\rho) \theta
                \end{aligned}
                \end{algomathdisplay}
                }
            }
            	Save current PID parameters $\theta$\;
        }
    }
}
\end{algorithm}
\end{minipage}
}

The optimization of $J$ in line \eqref{eq:PolicyGradient_Iteration} relies on knowledge of the $Q$-function \eqref{eq:Qfunc}. The critic network $Q_{\phi_i}$ is iteratively approximated using a deep neural network with training data from replay memory (RM). RM is a fixed-size collection of tuples of the form $(s, u, s', r)$. The PID controller (actor ``network'') is given by \eqref{eq:SS2} and denoted as $\mu$, and the critic is $Q_{\phi_i}$. More concretely, we utilize a combination of a feedforward neural network and a recurrent neural network (RNN) for the critic. As discussed in \cref{subsec:PIDRL}, the state may contain past output information. The state vector passes through a specialized RNN called a ``gated recurrent unit'' (GRU) \citep{cho2014LearningPhrase}; the hidden state of the GRU is then passed as an input with the action through a DNN.\par
\begin{table}[tbh]
\begin{center}
\resizebox{\textwidth}{!}{%
\begin{tabular}{l* {2}{c}}
\toprule
Hyperparameter &    Symbol &    Nominal value\\
\midrule
Actor learning rate & $\alpha_c$ & $0.002$ \\
Critic learning rate & $\alpha_c$ & $0.002$ \\
Discount factor & $\gamma$ & $0.99$ \\
Target network update rate & $\rho$ & $0.995$ \\
Policy update delay &  \texttt{policy\_{delay}} & $2$ \\
Batch size & $\abs{\mathcal{B}}$ & $256$ \\
Replay buffer size & $\abs{\mathcal{D}_{\text{RM}}}$ & $10,000$ \\
Target noise & $\sigma$ & $0.2$ \\
\bottomrule
\end{tabular}
}

\end{center}
\caption{The recommended hyperparameter settings from this work.}
\label{table:table_hyperparameters}
\end{table}
\cref{alg:ActorCriticProposed} summarizes the training procedure and \cref{table:table_hyperparameters} gives the hyperparameter names, symbols, and settings. The reader is referred to \citet{fujimoto2018AddressingFunction} and the references therein for more details and background on the TD3 algorithm. The primary hyperparameters we modified were the learning rates (for the actor and critic), the critic network, and the policy update delay. The other hyperparameters were set to the default values suggested by \citet{achiam2018SpinningDeep}. The learning rates and policy update delay did not deviate much from their original values either. The main design choice was with the critic network.\par
In \cref{alg:ActorCriticProposed}, note the distinctions between the measurement dataset $\mathcal{D}_{\text{process}}$ and replay memory $\mathcal{D}_{\text{RM}}$, and $\mu_{\theta}$ and $\mu_{\text{PLC}}$. $\mathcal{D}_{\text{process}}$ is a dataset storing measurements from the process, such as, setpoints, process variables, control variables. When it is time to update the PID parameters, new measurements in $\mathcal{D}_{\text{process}}$ are converted to the structured transition tuples $(s, u, s', r)$ the RL algorithm expects, and store them in $\mathcal{D}_{\text{RM}}$ for training. $\mu_{\theta}$ is defined as the actor ``network'' in our RL code, whereas $\mu_{\text{PLC}}$ uses the same parameters as $\mu_{\theta}$, but actually interacts with the system through a PLC. For the optimization of the actor and critic parameters, we use the Adam optimizer \citep{kingma2014AdamMethod} to implement the nominal update steps shown in \cref{line:critic_update} and \cref{line:actor_update}. \par

\section{Further experiments and discussion}
\label{app:experiments}

\begin{table*}[thb]
\begin{center}
     \resizebox{\linewidth}{!}{%
\begin{tabular}{ l *{16}{c} }
\toprule
\hspace*{2cm} &
\multicolumn{7}{c}{RL (Unconstrained)} &
\multicolumn{3}{c}{RL (Constrained)} & \multicolumn{5}{c}{SIMC} & \multicolumn{1}{c}{Accutune}\\
\cmidrule(lr){2-8} \cmidrule(lr){9-11} \cmidrule(lr){12-16} \cmidrule(lr){17-17}
&
\makebox[1em]{1} &
\textbf{\makebox[1em]{2}} &
\textbf{\makebox[1em]{3}} &
\makebox[1em]{4} &
\makebox[1em]{5} &
\makebox[1em]{6} &
\makebox[1em]{7} &
\textbf{\makebox[1em]{8}} &
\makebox[1em]{9} &
\makebox[1em]{10} &
\makebox[1em]{1} &
\makebox[1em]{2} &
\textbf{\makebox[1em]{3}} &
\makebox[1em]{4} &
\makebox[1em]{5} &
\textbf{\makebox[1em]{1}} \\
\midrule
Reward & Eq. \eqref{eq:baselineReward1} & \eqref{eq:baselineReward2} & \eqref{eq:baselineReward2} & \eqref{eq:baselineReward1} & \eqref{eq:hybridPenalty} & \eqref{eq:hybridReward} & \eqref{eq:hybridReward} & \eqref{eq:baselineReward2} & \eqref{eq:hybridReward} & \eqref{eq:hybridReward} 	& - & - & - & - & - & - 			\\
Initialization & $k_p=$ 4.0 & 4.0 & 2.0 & 2.0 & 4.0 & 4.0 & 2.0 & 4.0 & 4.0 & 4.0 	& - & - & - & - & - & - 				\\
Setting & - & - & - & - & - & - & - & - & - & - & $T_c=$ 9.21 & 15 & 20 & 25 & 30 & 	Mode=``slow"\\
\bottomrule
\end{tabular}
}\label{table:RLexperimentSpecs}
     \end{center}
\caption{Labels and specifications for RL, SIMC, and Accutune III experiments. In this paper we refer to them, for example, as ``RL-1'' for the first experiment in the RL category. The initialization refers to \cref{eq:initPID}.}
\label{table:experimentSpecs}
\end{table*}
\begin{figure}[!htb]
\begin{center}
\includegraphics[width=\linewidth]{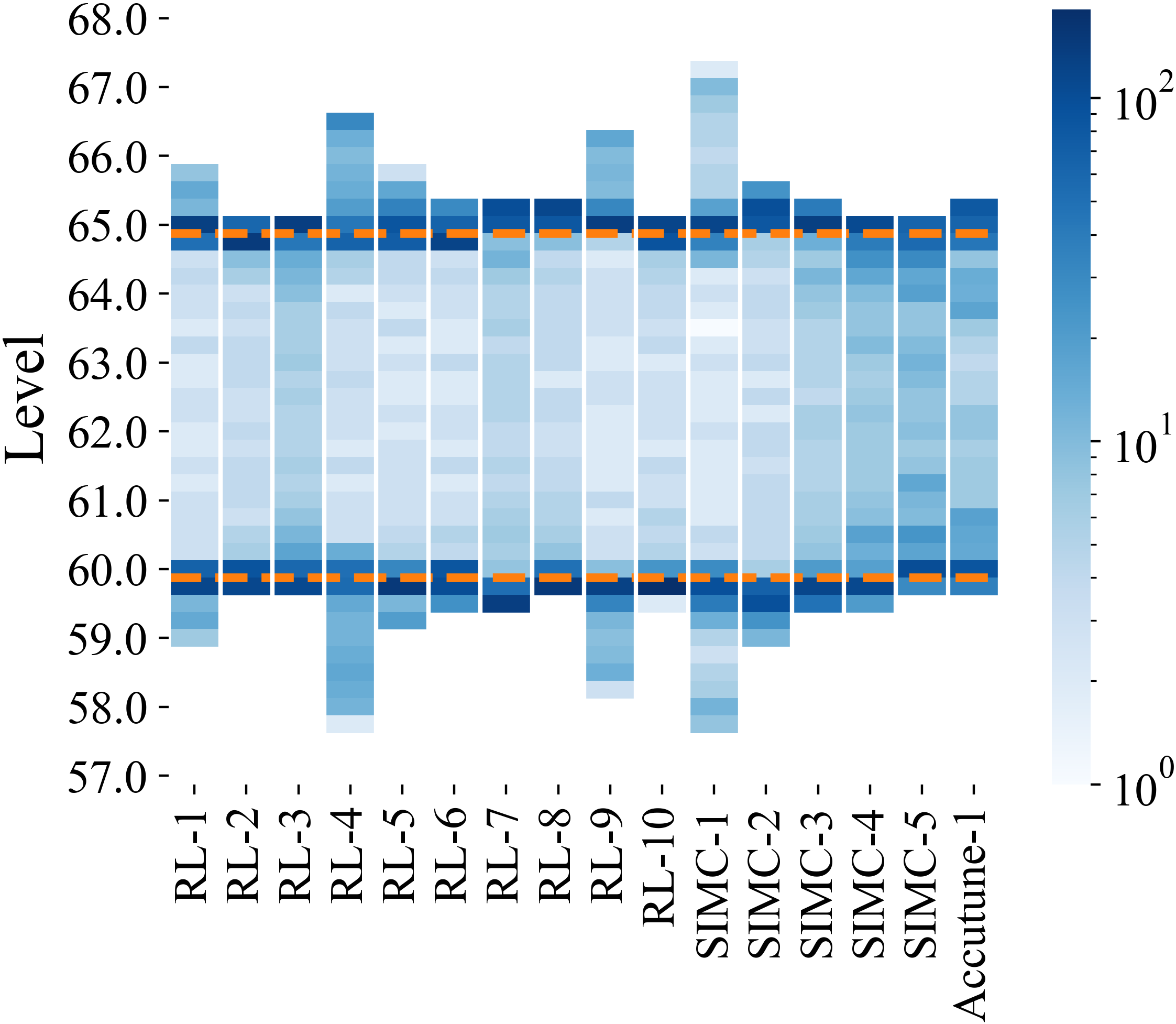}
\end{center}
\caption{Heatmap of final performances across many different experiments. Darker colors indicate more time spent by the process variable at that location on the y-axis. Dashed lines at $60$~cm and $65$~cm indicate setpoints.}
\label{fig:heatmap_allExperiments_nominal}
\end{figure}
We provide additional experiments and analysis. All the lab results are visualized in \cref{fig:heatmap_allExperiments_nominal} and performance statistics are reported in \cref{table:table_Robustness_allExperiments}. The experimental specifications corresponding to the labels given in these figures and tables are listed in \cref{table:experimentSpecs}. \rowref{RL}{row:robustness_td3_2021_04_07_1208}, \rowref{RL}{row:robustness_td3_2021_04_07_1357}, \rowref{RL}{row:robustness_td3_2021_04_28_1451}, \rowref{SIMC}{row:robustness_SIMC20}, \rowref{Accutune}{row:robustness_AccuSlow}, shown in boldface, refer to the five experiments shown in \cref{sec:results}. The difference between \rowref{RL}{row:robustness_td3_2021_05_14_1127} and \rowref{RL}{row:robustness_td3_2021_07_12_1047} is the transient training data: \rowref{RL}{row:robustness_td3_2021_05_14_1127} only switches between the setpoints $60$~cm and $65$~cm, where it hits the constraints at each step change.\par
We evaluate the performance on a variety of different reward functions based on the discussion in \cref{subsec:reward}. In addition to \cref{eq:baselineReward2} and the hybrid cost function $l$ given in \cref{eq:hybridReward}, we choose the reward functions:
\begin{align}
-r(s_t, u_t) &= e_t^2 + 0.1 \Delta u_t^2\label{eq:baselineReward1}\\
-r(s_t, u_t) &= l(s_t, u_t) + 0.1 (\Delta u_t)^2.\label{eq:hybridPenalty}
\end{align}
The rationale for comparing the end training results of four different reward functions is to demonstrate their relative performance according to the metrics in \cref{eq:metrics}.\par
\cref{fig:heatmap_allExperiments_nominal} shows a heatmap of the output performance of all the final tuning parameters from our experiments. Darker shades correspond to more time spent by the process variable at a value on the $y$-axis. This is a compact way of visualizing many experiments side-by-side while qualitatively capturing useful information such as overshoot or rise/settling time. Compared to the RL experiments involving the squared error based reward function (\rowref{RL}{row:robustness_td3_2021_04_07_1103},\rowref{RL}{row:robustness_td3_2021_04_07_1518}), the absolute value based reward performs better and is more consistent between training experiments. The hybrid reward function (\rowref{RL}{row:robustness_td3_2021_04_29_1040}, \rowref{RL}{row:robustness_td3_2021_04_29_1133}, \rowref{RL}{row:robustness_td3_2021_04_30_1426}, yields similar, and sometimes superior, nominal performance to the absolute value based reward.  

\subsection{Experimental robustness}
\label{appsubsec:robustness}

\begin{table*}[tbh]
\begin{center}
\resizebox{\linewidth}{!}{%
\begin{tabular}{NPPPPPPc}
\toprule
\multicolumn{0}{l}{}  &              \multicolumn{2}{r}{IAE} &              \multicolumn{2}{c}{ISE} &            \multicolumn{2}{c}{TV} &          \multicolumn{2}{c}{TV$_u$} &          \multicolumn{2}{c}{$\%$ OS} &                \multicolumn{2}{c}{ST} & $\text{M}_s$ \\
\midrule
RL-\therowcntr\label{row:robustness_td3_2021_04_07_1103}   &   44.41&9.57 &   28.33&4.89 &  1.88&0.23 &  11.14&1.08 &  22.94&10.77 &  134.67&25.51 & 1.55\\
\textbf{RL-\therowcntr}\label{row:robustness_td3_2021_04_07_1208}   &   43.76&7.59 &   30.18&3.85 &  1.56&0.12 &   9.63&0.50 &   10.95&8.26 &  129.08&34.22 & 1.35\\
\textbf{RL-\therowcntr}\label{row:robustness_td3_2021_04_07_1357}   &   48.75&4.55 &   34.00&3.40 &  1.42&0.05 &   8.90&0.20 &    7.46&2.03 &  153.17&26.34 & 1.25\\
RL-\therowcntr\label{row:robustness_td3_2021_04_07_1518}   &  67.96&15.42 &  40.09&10.96 &  2.32&0.40 &  12.81&1.92 &  48.20&14.38 &  186.58&28.46 & 1.42\\
RL-\therowcntr\label{row:robustness_td3_2021_04_29_1040}   &   42.72&8.03 &   27.87&4.16 &  1.78&0.21 &  10.67&0.91 &   20.43&9.74 &  128.75&15.36 & 1.48\\
RL-\therowcntr\label{row:robustness_td3_2021_04_29_1133}   &   40.85&7.63 &   27.00&3.81 &  1.69&0.23 &  10.32&1.14 &   17.33&9.84 &  145.42&41.48 & 1.50\\
RL-\therowcntr\label{row:robustness_td3_2021_04_30_1426}       &   54.92&6.65 &   34.65&2.94 &  1.50&0.09 &      9.06&0.47 &   14.87&4.67 &  190.58&34.84 & 1.27\\
\textbf{RL-\therowcntr}\label{row:robustness_td3_2021_04_28_1451}   &   42.75&5.87 &   28.89&3.75 &  1.56&0.09 &   9.62&0.41 &   11.38&6.48 &  166.42&20.03 & 1.36\\
RL-\therowcntr\label{row:robustness_td3_2021_05_14_1127}   &  56.47&13.25 &   33.90&7.48 &  2.06&0.25 &  11.45&1.25 &   37.84&9.84 &  163.08&21.80 & 1.51\\
RL-\therowcntr\label{row:robustness_td3_2021_07_12_1047}\setcounter{rowcntr}{0}  &   40.18&6.59 &   27.84&3.61 &  1.58&0.13 &   9.81&0.52 &   11.08&7.57 &   133.17&5.63 & 1.41\\
SIMC-\therowcntr\label{row:robustness_SIMC9} &  61.64&17.36 &  36.73&10.19 &  2.68&0.72 &  14.29&3.40 &  53.55&14.48 &  162.08&52.19 & 1.72\\
SIMC-\therowcntr\label{row:robustness_SIMC15} &  56.16&10.89 &   34.36&6.05 &  1.73&0.11 &  10.07&0.49 &   24.11&7.83 &  168.17&29.17 & 1.36\\
\textbf{SIMC-\therowcntr}\label{row:robustness_SIMC20} &   54.74&6.49 &   35.47&3.20 &  1.52&0.02 &   9.17&0.05 &   13.26&4.97 &  193.75&19.16 & 1.25\\
SIMC-\therowcntr\label{row:robustness_SIMC25} &   57.61&5.08 &   39.12&3.50 &  1.38&0.02 &   8.49&0.11 &    8.38&2.57 &  167.17&47.65 & 1.19\\
SIMC-\therowcntr\label{row:robustness_SIMC30}\setcounter{rowcntr}{0} &  57.33&12.18 &   38.77&9.85 &  1.33&0.06 &   8.31&0.34 &    4.86&1.79 &  141.67&13.73 & 1.16\\
\textbf{Accutune-\therowcntr}\label{row:robustness_AccuSlow}  &  70.09&24.46 &  46.33&16.94 &  1.59&0.24 &  351.62&201.09 &    8.98&0.46 &  139.33&65.07 & 1.41\\
\bottomrule
\end{tabular}
}

\end{center}
\caption{The overall performance of each experiment. Each cell shows the average statistic for its column plus or minus the standard deviation; the average is computed over the nominal performance and that of two robustness experiments where we independently adjusted the outflow of the tank and the flow tuning parameters.}
\label{table:table_Robustness_allExperiments}
\end{table*}
So far we have reported on the promising nominal performance of the RL based tuning. We also evaluate the empirical robustness of these results. With each set of final tuning parameters from our experiments, we perform the same sequence of step changes reported in \cref{sec:results} ($65$~cm, $60$~cm, $63$~cm, $60$~cm), but with the following independent changes to the two-tank system: First, we tighten the outflow from the top tank to {$50\%$} on its valve (whereas before it was at maximum outflow); then, with the outflow back at its original state, we detune the flow controller. The flow controller parameters were originally set to ${k_p = 0.2}$, ${k_i = k_p/3}$, ${k_d = 0.67k_p}$, ${T_f = 0.1}$; we cut $k_p$ in half for the second experiment.\par
Results for these experiments are reported in \cref{table:table_Robustness_allExperiments}. It encompasses the nominal performance and the performance of the two aforementioned robustness experiments. For each statistic, performance is measured based on the average across the four step changes. Each cell shows the average performance across these three experiments plus or minus the standard deviation.\par 
Out of \rowref{RL}{row:robustness_td3_2021_04_07_1103} -- \rowref{RL}{row:robustness_td3_2021_04_07_1518}, \rowref{RL}{row:robustness_td3_2021_04_07_1208} and \rowref{RL}{row:robustness_td3_2021_04_07_1357} remain the most promising results. With the exception of ISE between \rowref{RL}{row:robustness_td3_2021_04_07_1103} and \rowref{RL}{row:robustness_td3_2021_04_07_1208}, these two experiments based on the the reward function from \cref{eq:baselineReward2} are superior to \rowref{RL}{row:robustness_td3_2021_04_07_1103} and \rowref{RL}{row:robustness_td3_2021_04_07_1518}, respectively, across all categories, both in terms of their average and standard deviation.\par 
For the hybrid reward function, given in \cref{eq:hybridReward}, it is apparent that \rowref{RL}{row:robustness_td3_2021_04_29_1133} gives the best IAE and ISE, but with worse overshoot than \rowref{RL}{row:robustness_td3_2021_04_07_1208}. Moreover, \rowref{RL}{row:robustness_td3_2021_04_30_1426} indicates a wider detrimental change in IAE and ISE based on the initial tuning than that of \rowref{RL}{row:robustness_td3_2021_04_07_1208} and \rowref{RL}{row:robustness_td3_2021_04_07_1357}. For the constrained experiments, \rowref{RL}{row:robustness_td3_2021_04_28_1451} and \rowref{RL}{row:robustness_td3_2021_07_12_1047} achieved similar scores for TV, $\text{TV}_u$, and $\% \text{OS}$, with \rowref{RL}{row:robustness_td3_2021_07_12_1047} slightly superior in terms of IAE, ISE, and ST. Despite this, we consider the experiments based on the reward in \cref{eq:baselineReward2} to be the overall best option; this is in terms of performance, generalization between training experiments, and robustness.\par


%

\end{document}